\documentclass[times,twocolumn,final]{elsarticle}

\usepackage{medima}
\usepackage{framed,multirow}
\usepackage{amssymb}
\usepackage{latexsym}
\usepackage{array}
\usepackage{amsmath}
\usepackage{textcomp}
\usepackage{url}
\usepackage[table,xcdraw]{xcolor}
\usepackage[hidelinks]{hyperref}
\usepackage{float}

\begin{document}

\verso{Schuler \textit{et~al.}: Consistent biventricular coordinates}

\begin{frontmatter}

\title{Cobiveco: Consistent biventricular coordinates for precise and intuitive description of position in the heart -- with MATLAB implementation}

\fntext[fn1]{These authors contributed equally.}
\author[1]{Steffen \snm{Schuler}\corref{cor1}\fnref{fn1}}
\cortext[cor1]{Corresponding author: 
Institute of Biomedical Engineering,
Karlsruhe Institute of Technology (KIT),
Fritz-Haber-Weg 1,
76131 Karlsruhe,
Germany.
Email address: publications@ibt.kit.edu.}
\author[1]{Nicolas \snm{Pilia}\fnref{fn1}}
\author[2]{Danila \snm{Potyagaylo}}
\author[1]{Axel \snm{Loewe}}
\address[1]{Institute of Biomedical Engineering, Karlsruhe Institute of Technology (KIT), 76131 Karlsruhe, Germany}
\address[2]{EPIQure GmbH, 76139 Karlsruhe, Germany}

\begin{abstract}
Ventricular coordinates are widely used as a versatile tool for various applications that benefit from a description of local position within the heart.
However, the practical usefulness of ventricular coordinates is determined by their ability to meet application-specific requirements.
For regression-based estimation of biventricular position, for example, a symmetric definition of coordinate directions in both ventricles is important. For the transfer of data between different hearts as another use case, the consistency of coordinate values across different geometries is particularly relevant. To meet these requirements, we compare different approaches to compute coordinates and present Cobiveco, a symmetric, consistent and intuitive biventricular coordinate system that builds upon existing coordinate systems, but overcomes some of their limitations. A novel one-way transfer error is introduced to assess the consistency of the coordinates. Normalized distances along bijective trajectories between two boundaries were found to be superior to solutions of Laplace's equation for defining coordinate values, as they show better linearity in space. Evaluation of transfer and linearity errors on 36 patient geometries revealed a more than 4-fold improvement compared to a state-of-the-art method. Finally, we show two application examples underlining the relevance for cardiac data processing. Cobiveco MATLAB code is available under a permissive open-source license.
\\\ \\
{\footnotesize This article has been published in Medical Image Analysis (\href{https://doi.org/10.1016/j.media.2021.102247}{https://doi.org/10.1016/j.media.2021.102247}).}
\end{abstract}


\end{frontmatter}

%%%%%%%%%%%%%%%%%%%%%%%%%%%%%%%%%%%%%%%%%%%%%%%%%%
\section{Introduction}
\label{introduction}
Patient and species independent representations of ventricular anatomy are a valuable tool for data processing in cardiology. Typical applications include the standardized visualization and regional evaluation of cardiac data, a transfer of data between different hearts from different measurement modalities, and the description of local position in the heart. Recently, such representations have become particularly important for non-invasive localization of the excitation origin using machine learning algorithms~\citep{Yang-2018-ID12792,Zhou-2019-ID13157}.\\
The most popular example of such a representation is the AHA segmentation from~\cite{cerqueira01} which divides the left ventricle (LV) into 17 segments. While easy to apply in practice, it only allows a discrete, coarse-grained representation of the LV and does not cover the right ventricle (RV).\\
The approach proposed by~\cite{Paun-2017-ID11710} is more general as it provides a continuous parameterization of both LV and RV. It uses solutions to Laplace's equation to flatten a ventricular bounding surface onto a planar domain and to encode the thickness of anatomical structures on top of this planar domain. Although intended for detailed representation of the ventricular interior (endocardium, trabeculations, papillary muscles), it may also be applied to the whole myocardial wall. However, this approach does not directly provide an intuitive description of local position and treats LV and RV independently.\\
The universal ventricular coordinates (UVC) introduced by~\cite{Bayer-2018-ID11708} offer such an intuitive description by defining an apicobasal, a rotational, a transmural and a transventricular coordinate -- each of which is defined using solutions to Laplace's equation. UVC thereby offer a parameterized description of ventricular position and a similar method exists for the atria~\citep{Roney-2019-ID14879}. Although clearly an excellent idea, the UVC system may be improved in three respects: First, the definition of coordinates in the LV and RV is not symmetric, which causes discontinuities at the junctions of LV and RV and leads to holes in the coordinate space. This can be problematic, for instance, for regression-based estimation of ventricular position. Second, the apicobasal coordinate must go to zero at singularities of the rotational coordinate, in order to make these locations uniquely characterized by the coordinates. This is not guaranteed by the construction of boundary conditions for the apicobasal and rotational coordinates. Third, the consistency of coordinate values across different geometries is in certain cases not optimal, as solutions to Laplace's equation are in general not an accurate measure of (normalized) distance. These aspects can lead to errors in transferring data between hearts and other applications that rely on a consistent description of local position. This especially pertains to the validation of electrocardiographic imaging~\citep{Cluitmans-2018-ID12162}. Here, the transfer of potentials or activation times from a geometry obtained using intracardiac mapping onto a tomography-derived geometry, which is used for inverse reconstructions, is usually needed. Inconsistencies between coordinates computed on these two geometries can cause problematic artifacts in the transferred signals.\\
In this work, we propose a coordinate system for biventricular geometries that builds upon the pioneering work by~\cite{Bayer-2018-ID11708} and previous works but reduces transfer errors. We start by defining desirable properties for such a coordinate system. Then we explain the underlying concept for symmetric and consistent coordinate directions and compare different approaches for computing the actual coordinate values based on partial differential equations (PDE). Having identified a suitable approach, we provide a detailed description of the new coordinate system, called \textit{Cobiveco}. Finally, we compare Cobiveco with UVC by evaluating transfer and linearity errors and present application examples.

%%%%%%%%%%%%%%%%%%%%%%%%%%%%%%%%%%%%%%%%%%%%%%%%%%
\section{Methods}
\label{methods}

\subsection{Desirable properties for biventricular coordinates}
\label{properties}
Based on the use cases mentioned in the introduction, the following properties are considered desirable for a biventricular coordinate system:
\begin{itemize}
	\setlength\itemsep{0em}
	\item \textit{Bijective:} Each coordinate tuple corresponds to exactly one point in the heart.
	\item \textit{Continuous:} Coordinates have no jumps.
	\item \textit{Normalized:} Coordinates range between 0 and 1.
	\item \textit{Complete:} Each tuple in this range represents a valid position, i.e., the coordinate space has no holes.
	\item \textit{Linear:} Coordinates change linearly in space, i.e., the geodesic distance traveled when changing one coordinate, while keeping all others fixed, is proportional to the change in this coordinate.
	\item \textit{Symmetric parameterization:} The underlying parameterization is the same for both ventricles.
	\item \textit{Consistent landmarks:} Clear anatomical landmarks are represented by the same coordinates across different hearts. In particular, landmarks used to construct the coordinate system are robust to variations in shape.
\end{itemize}

Note that normalized and linear coordinates can in general (for arbitrary shapes) not also be orthogonal\footnote{For example, shearing a rectangular domain on which normalized cartesian coordinates are defined destroys the orthogonality of the coordinates, while preserving their linearity.}. The resulting coordinate system will not preserve angles, but it will preserve distances in each of the coordinate directions.

\subsection{Concept for symmetric coordinate directions}
\label{concept}
\begin{figure}[H]
	\centering
	\includegraphics[width=\linewidth]{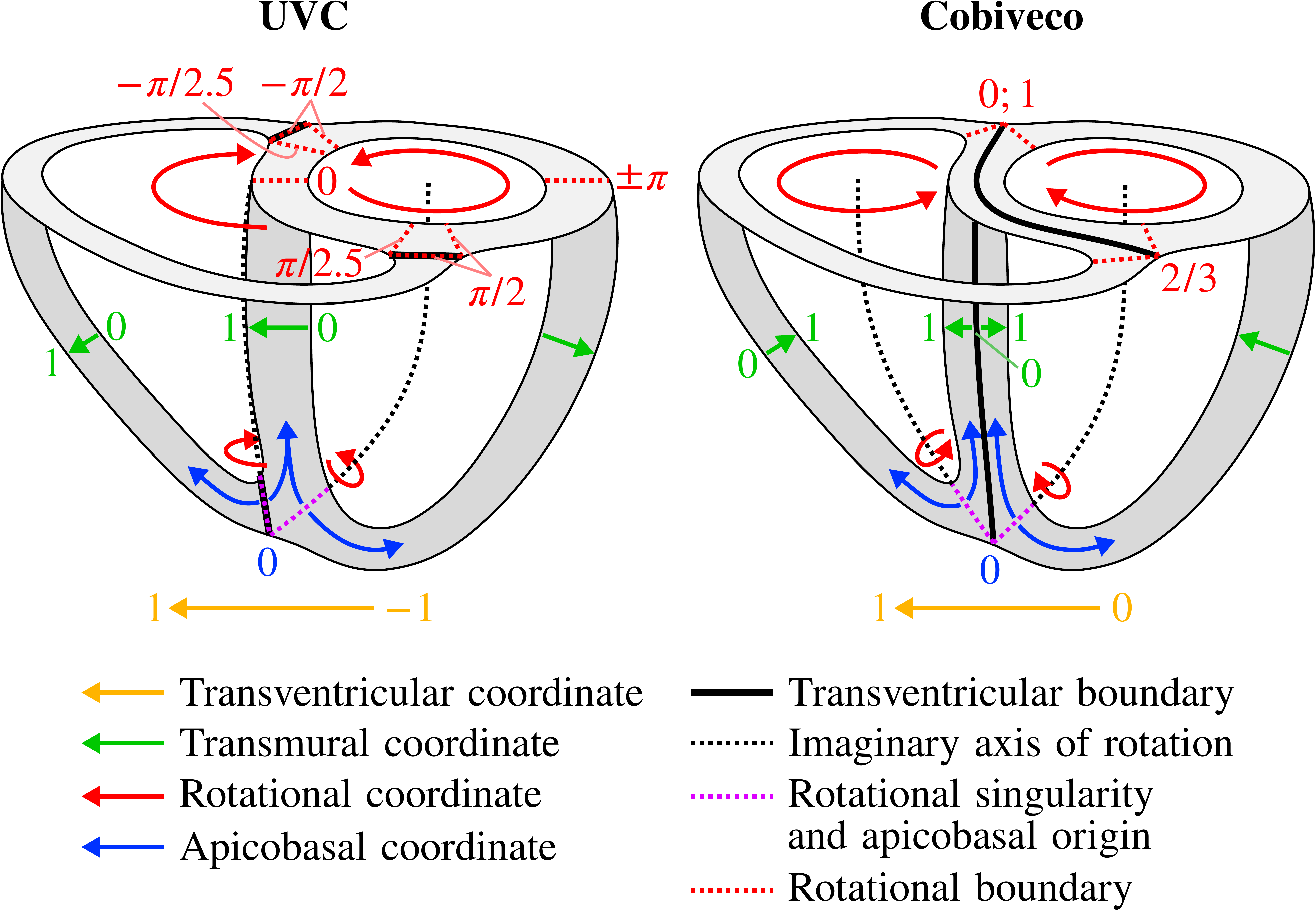}
	\caption{Different underlying concepts for coordinate directions and boundary values within the UVC (\textit{left}) and the suggested coordinate system Cobiveco (\textit{right}). A basal cross-section in long-axis direction and a central cross-section in anterior-posterior direction are shown.}
	\label{fig:concept}
\end{figure}

Following the UVC approach from~\cite{Bayer-2018-ID11708}, our choice of coordinate directions is inspired by prolate spheroidal coordinates as used in~\cite{costa96} to parameterize an idealized LV geometry using an apicobasal, a rotational and a transmural coordinate. The goal of UVC and Cobiveco is to find a ``generalization'' for biventricular geometries of arbitrary shape. To this end, one more transventricular coordinate is needed that distinguishes between LV and RV.\\
The left panel of Fig.~\ref{fig:concept} illustrates the basic concept for these four coordinates within the UVC system. Here, the transventricular boundary is chosen such that the entire septum belongs to the LV. While this choice is anatomically intuitive and might be most useful for applications focusing on the LV, it leads to undesired properties of the coordinates: The transmural and the rotational coordinates are discontinuous at the transventricular boundary and the ranges of the rotational coordinate are different in the LV and the RV ($-\pi$ to $\pi$ vs. $-\pi/2$ to $\pi/2$).\\
To overcome these inconsistencies, we suggest to move the transventricular boundary to the center of the septum, as shown in the right panel of Fig.~\ref{fig:concept}. This results in entirely symmetric transmural, rotational and apicobasal coordinates in both ventricles and removes discontinuities. The transmural coordinate increases from the center of the septum, so that both sides of the septal endocardium have the same value. The rotational coordinate is counter-rotating in the LV and RV free walls and unifies at the septum. It is normalized to also range from $0$ to $1$.

\subsection{Comparison of PDE-based approaches to compute coordinate values}
\label{comparisonOfApproaches}
Solving PDEs can be an efficient and elegant way to compute coordinate values. However, the type of PDE and the boundary conditions need to be chosen with care. In this section, different PDE-based approaches to compute coordinate values are presented and compared to each other to choose the most adequate one.
Solutions to Laplace's equation (in the following just ``Laplace solutions'') in between two boundaries with Dirichlet conditions are one obvious approach and were utilized in~\cite{Bayer-2018-ID11708}. Nevertheless, they are not necessarily a good choice as their linearity severely depends on the width of the domain between these boundaries.
This follows directly from the divergence theorem: As the Laplace equation requires the divergence of the gradient to be zero and the flux of the gradient field through lateral parts of the outer surface is already zero due to zero Neumann boundary conditions, the (signed) fluxes of the gradient field through any two cross-sectional surfaces have to compensate each other.
This implies that smaller gradients occur in wider regions and vice versa, which makes Laplace solutions an unreliable measure of normalized distance between boundaries. Therefore, they should not directly be used to define coordinate values.
In the UVC method, this becomes apparent for the apicobasal Laplace solution in between a small apical and a large basal boundary, where it led the UVC authors to normalize the resulting apicobasal coordinate using the values on the shortest geodesic path between apex and base~\citep{Bayer-2018-ID11708}. However, it can also lead to substantial distortions of the transmural and rotational coordinates.

\begin{figure}[t]
	\centering
	\includegraphics[width=\linewidth]{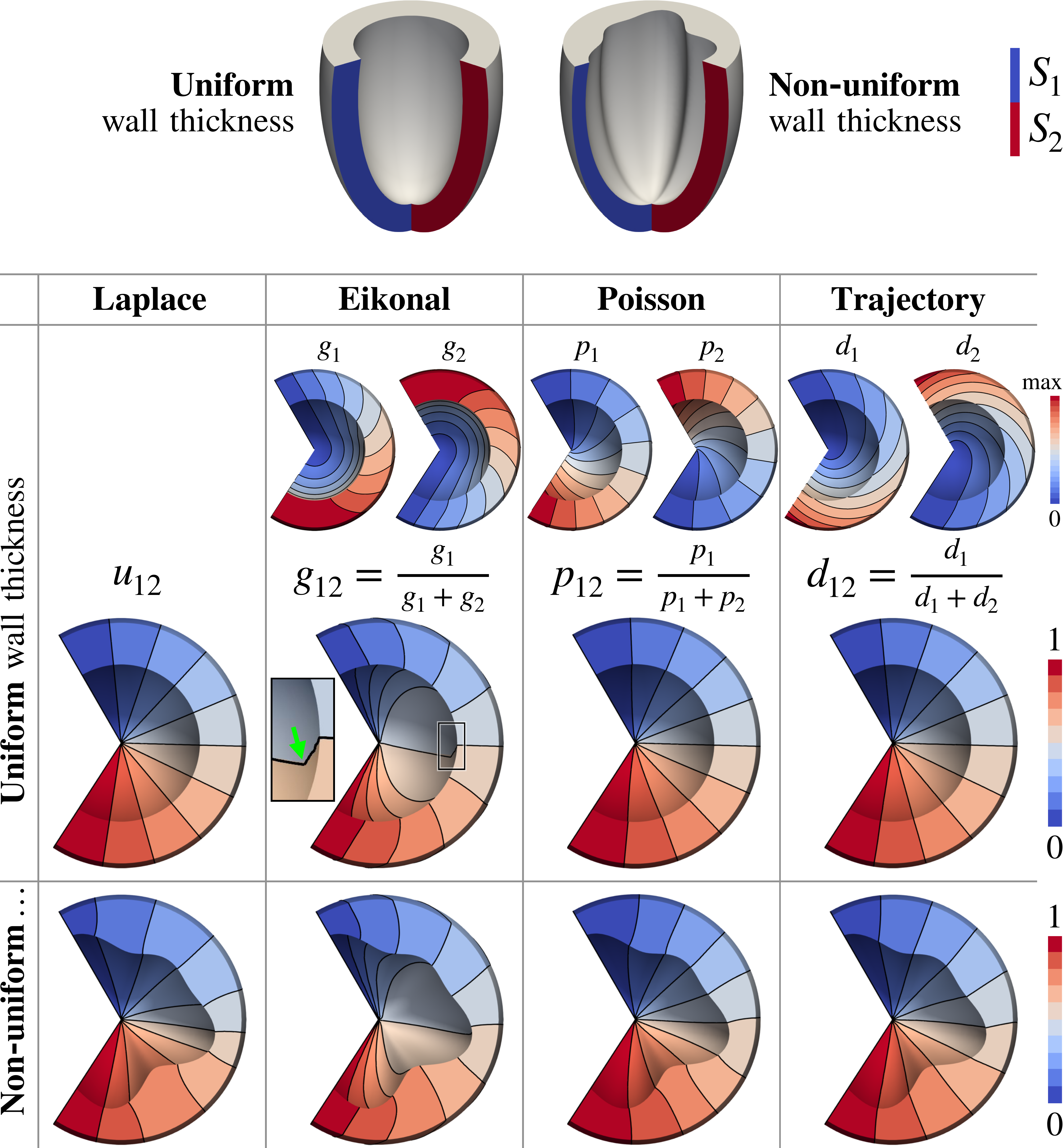}
	\caption{Comparison of four approaches to compute a rotational coordinate. \textit{Top:} Ellipsoidal geometries with boundary surfaces used as input. \textit{Bottom:} Results for the different approaches. For the Eikonal approach, several cusps occur at the lateral surfaces. The green arrow marks one of them. The Eikonal solutions were computed using the fast iterative method \citep{Fu-2013-ID14300}, while all other solutions were computed in MATLAB (see section \ref{novelCoords}).}
	\label{fig:approaches}
\end{figure}

To demonstrate the effect on the rotational coordinate, we created two ellipsoidal geometries resembling the LV free wall: One with uniform and one with non-uniform wall thickness as expected in reality. The two geometries with the boundary surfaces $S_1$ and $S_2$ are shown at the top of Fig.~\ref{fig:approaches}. The Laplace solution $u_{12}$ between these two boundary surfaces is depicted in the first column of Fig.~\ref{fig:approaches}:
\begin{equation}
	\Delta u_{12} = 0 \quad\text{with}\quad u_{12}(S_1)=0 \quad\text{and}\quad u_{12}(S_2)=1
\end{equation}
The result for the case of a uniform wall thickness is as desired, i.e., the values change linearly between the two boundaries. However, distortions can be seen for the non-uniform case. In this example, the rotational distance between contour lines is more than twice as large in the thickest region compared to the thinnest region, calling for a better approach.

Using the Eikonal equation instead of the Laplace equation might seem a natural choice to yield equidistant contour lines. But only non-normalized distances $g_1$ and $g_2$ with respect to a single boundary can be obtained:
\begin{alignat}{2}
	\|\nabla g_1\| &= 1 \quad\text{with}\quad g_1(S_1)\ &= 0\\
	\|\nabla g_2\| &= 1 \quad\text{with}\quad g_2(S_2)\ &= 0
\end{alignat}
A simple way to get a normalized ``distance'' between both boundaries is to compute the following quotient:
\begin{equation}
	g_{12} = \frac{g_1}{g_1+g_2}
	\label{eq:eikonalQuotient}
\end{equation}
However, the result shows a very inhomogeneous distribution of contour lines, even for the case of uniform wall thickness (second column of Fig.~\ref{fig:approaches}). Furthermore, the contour lines often have cusps (green arrow). The reason is that $g_1$ and $g_2$ represent distances along different, non-bijective trajectories between $S_1$ and $S_2$, which makes the normalization according to \eqref{eq:eikonalQuotient} invalid.

Another strategy to reduce the non-linearities of the Laplace solution is to compute its gradient field, normalize it to unit length and ``integrate it back'' by solving Poisson's equation:
\begin{equation}
	\mathbf{t}_{12} = \frac{\nabla u_{12}}{\|\nabla u_{12}\|}
\end{equation}
\vspace{-1.5\baselineskip}
\begin{alignat}{2}
	\Delta p_1 &= &\nabla\cdot \mathbf{t}_{12} \quad\text{with}\quad p_1(S_1) &= 0 \label{eq:poisson1}\\
	\Delta p_2 &= -&\nabla\cdot \mathbf{t}_{12} \quad\text{with}\quad p_2(S_2) &= 0 \label{eq:poisson2}
\end{alignat}
This approach is inspired by the heat method for computing geodesic distances~\citep{Crane-2013-ID13149}.    
However, this also yields non-normalized distances. Normalization as in \eqref{eq:eikonalQuotient} gives a satisfactory result $p_{12}$ for the uniform but not for the non-uniform case (third column of Fig.~\ref{fig:approaches}). The problem here is that although the trajectories along $\mathbf{t}_{12}$ are bijective, the trajectories along $\nabla p_1$ and $\nabla p_2$ are not anymore, due to the divergence operator in \eqref{eq:poisson1},\,\eqref{eq:poisson2}. Therefore, Poisson's equation is not an appropriate way to integrate $\mathbf{t}_{12}$ for our purpose.
A further way is by solving the ``trajectory distance equation'' originally proposed for obtaining a symmetric measure of tissue thickness by~\cite{Yezzi-2003-ID13374}:
\vspace{-1.5\baselineskip}
\begin{alignat}{2}
	\nabla d_1 \cdot \mathbf{t}_{12} &= 1 \quad\text{with}\quad d_1(S_1)\ &= 0\\
	-\nabla d_2 \cdot \mathbf{t}_{12} &= 1 \quad\text{with}\quad d_2(S_2)\ &= 0
\end{alignat}
These systems of linear equations are overdetermined, because there are more elements than nodes in a tetrahedral mesh. Hence, they are solved in a least-squares sense.
As the gradient fields of the trajectory distances $d_1$ and $d_2$ themselves now match $\mathbf{t}_{12}$ and $-\mathbf{t}_{12}$, respectively (and not only their divergences), they allow for normalization as in \eqref{eq:eikonalQuotient}. The normalized result $d_{12}$ is shown in the last column of Fig.~\ref{fig:approaches} and exhibits the desired behavior even for the non-uniform case.\\
We conclude that normalized distances obtained by solving the trajectory distance equation are well suited to define coordinate values and should be preferred over Laplace's, Poisson's or the Eikonal equation.

\subsection{New coordinate system ``Cobiveco''}
\label{novelCoords}
\begin{figure}[!b]
	\centering
	\includegraphics[width=\linewidth]{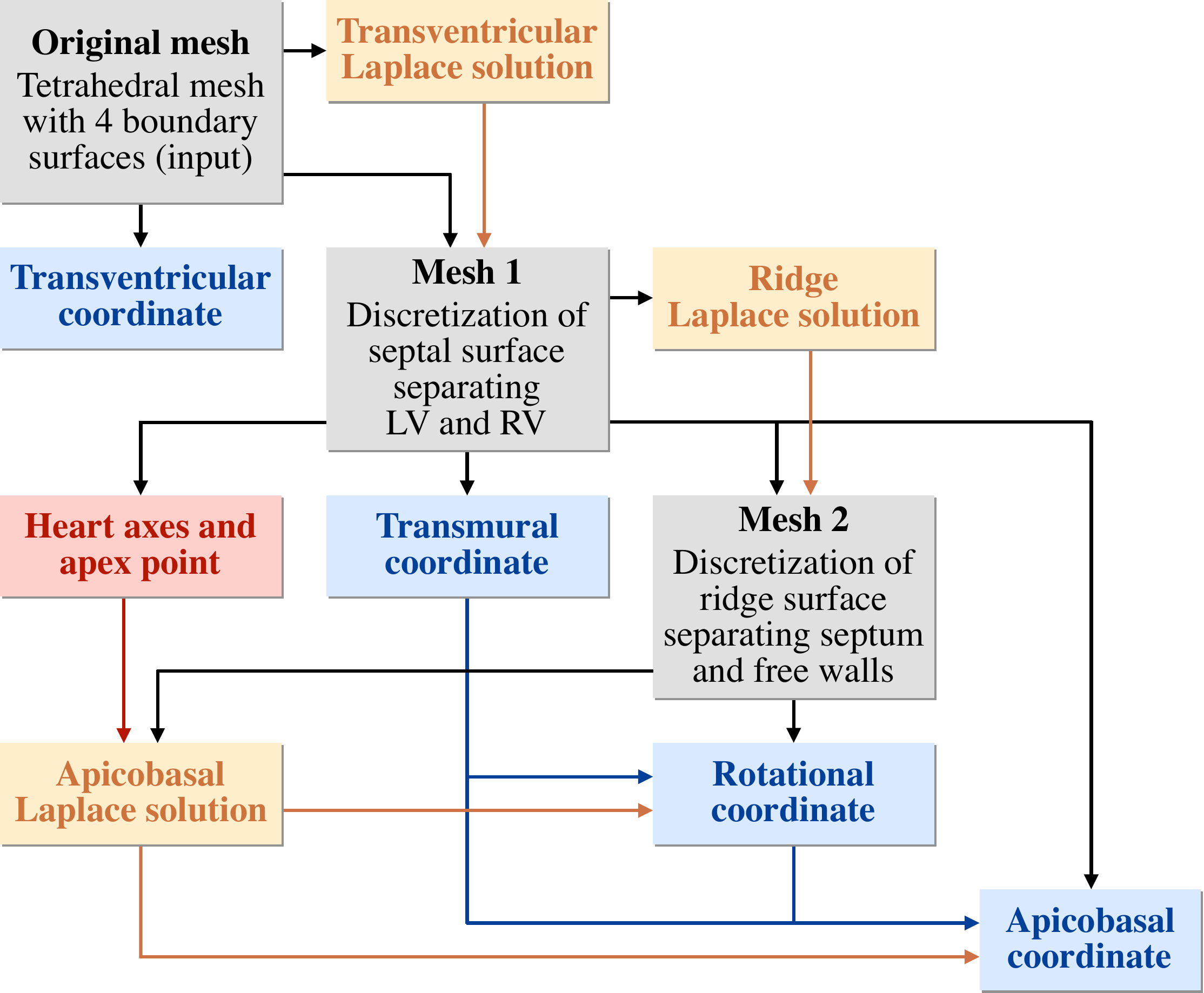}
	\vspace{-2mm}
	\caption{Overview of the process to compute Cobiveco, starting at the top-left and ending at the bottom-right. The gray, blue and red boxes represent steps of mesh preparation, actual coordinate computation and the extraction of additional axes and points, respectively, and each correspond to one of the sections \ref{inputs}--\ref{apicobasal}. The yellow boxes represent auxiliary Laplace solutions.}
	\label{fig:overview}
\end{figure}

In each of the following subsections, we will describe one of the eight steps involved in the computation of coordinate values according to Cobiveco. Fig.~\ref{fig:overview} provides an overview of the computational process. We use $V$ for denoting volumes, $S$ for surfaces, $C$ for curves and $\mathbf{x}$ for points. $u$ is used for Laplace solutions and $d$ for (relative) trajectory distances. The mathematical description is paralleled by a purely verbal description. To increase readability, we use verbal terms introduced in italic font instead of the corresponding mathematical symbols whenever possible in the text. The computational process on discrete meshes is described in the text and using pictures, while the mathematical notation refers to the continuous case.

We provide an open-source MATLAB implementation of \mbox{Cobiveco} for tetrahedral meshes (\url{https://github.com/KIT-IBT/Cobiveco}) under the Apache License 2.0. For more details about the implementation, the reader is referred to the code. To solve partial differential equations, we use the discrete Laplace and gradient operators from the \mbox{\textit{gptoolbox}}~\citep{Jacobson-2018-ID13392}, which are equivalent to first order finite element discretization~\citep{Jacobson-2013-ID13391}. For general geometry processing (thresholding, surface extraction, connectivity filtering, isocontour computation, etc.), the \textit{VTK library}~\citep{Schroeder-2006-8991} is used, for which we have developed and provide a MEX interface called \mbox{\textit{vtkToolbox}} (\url{https://github.com/KIT-IBT/vtkToolbox}). Implicit domain remeshing (isovalue discretization) is performed using \mbox{\textit{mmg3d}}~\citep{Dapogny-2014-ID13303}.

\subsubsection{Definition of inputs}
\label{inputs}
\begin{figure}[H]
	\centering
	\includegraphics[width=\linewidth]{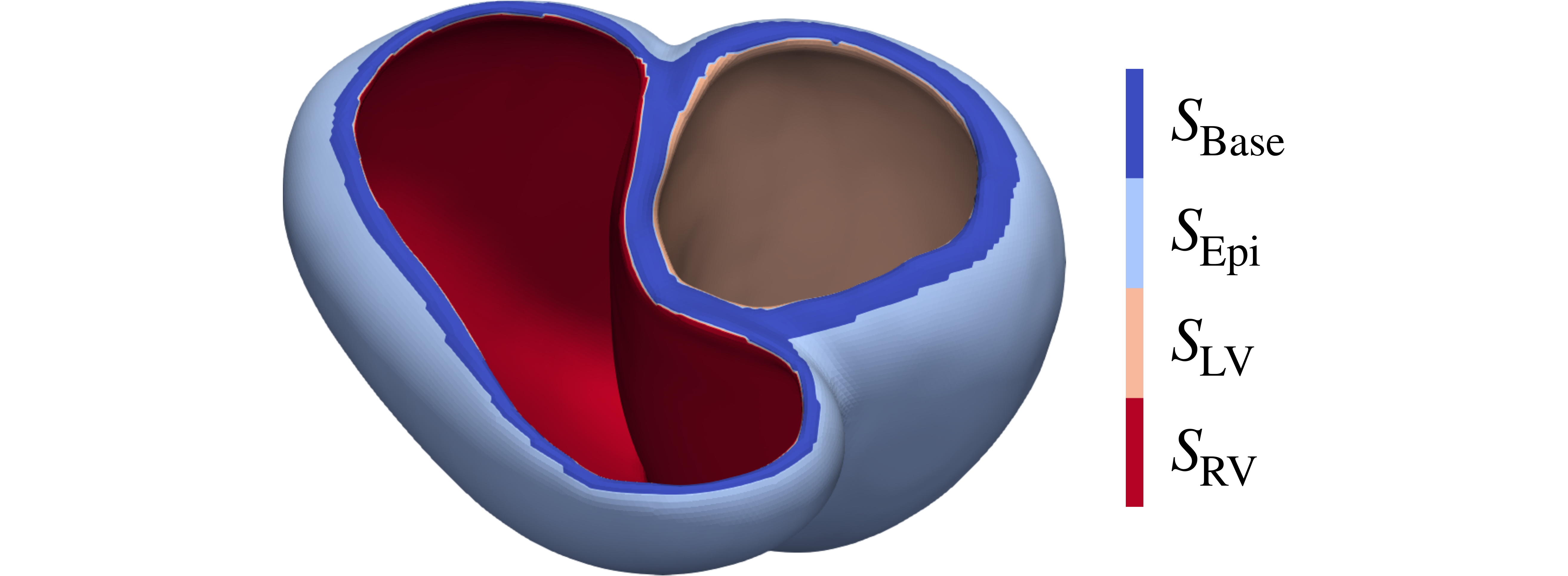}
	\caption{Boundary surfaces required as input: Basal surface, epicardial surface, LV endocardial surface and RV endocardial surface.}
	\label{fig:boundarySurfaces}
\end{figure}

Cobiveco requires a biventricular volume $V$ with exactly one orifice at the base of each ventricle. If there are bridges between the tricuspid valve and the RV outflow tract or between the mitral valve and the LV outflow tract, they have to be removed. To yield consistent results across different geometries, the base of the heart should be truncated at comparable heights.
Apart from the volume mesh, four boundary surfaces as depicted in Fig.~\ref{fig:boundarySurfaces} are needed as input: a \textit{basal surface}~$S_\mathrm{Base}$, an \textit{epicardial surface}~$S_\mathrm{Epi}$, an \textit{LV endocardial surface}~$S_\mathrm{LV}$, and an \textit{RV endocardial surface}~$S_\mathrm{RV}$.\\
We provide utilities for semi-automatic clipping at the base, removal of bridges and extraction of these boundary surfaces as part of the Cobiveco code.

\subsubsection{Computation of transventricular coordinate $v$}
\label{transventricular}
\begin{figure}[H]
	\vspace{-2mm}
	\centering
	\includegraphics[width=\linewidth]{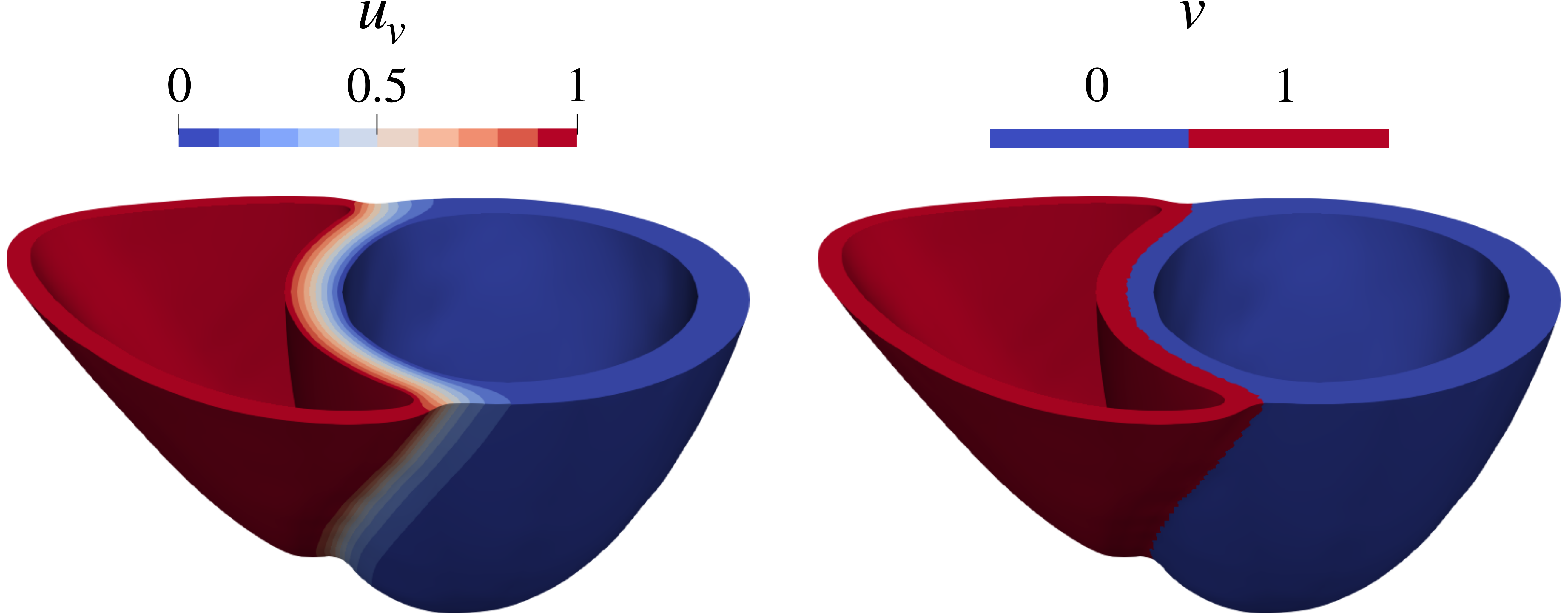}
	\caption{Computation of the transventricular coordinate. \textit{Left:} Laplace solution. \textit{Right:} Final coordinate. The geometry was clipped for visualization.}
	\label{fig:transventricular}
\end{figure}

To compute the transventricular coordinate, we first solve Laplace's equation with boundary conditions of 0 at the RV endocardium and 1 at the LV endocardium (Fig.~\ref{fig:transventricular}, left):
\begin{equation}
	\Delta u_v(V) = 0 \quad\text{with}\quad u_v(S_\mathrm{RV})=0 \quad\text{and}\quad u_v(S_\mathrm{LV})=1
	\label{eq:transventricularLap}
\end{equation}
This solution is then rounded, which yields the final \textit{transventricular coordinate} $v$ with binary values (Fig.~\ref{fig:transventricular}, right):
\begin{equation}
	v = \operatorname{round}(u_v)
\end{equation}

\subsubsection{First remeshing; extraction of septal surface and curve}
\label{mesh1}
\begin{figure}[H]
	\centering
	\includegraphics[width=\linewidth]{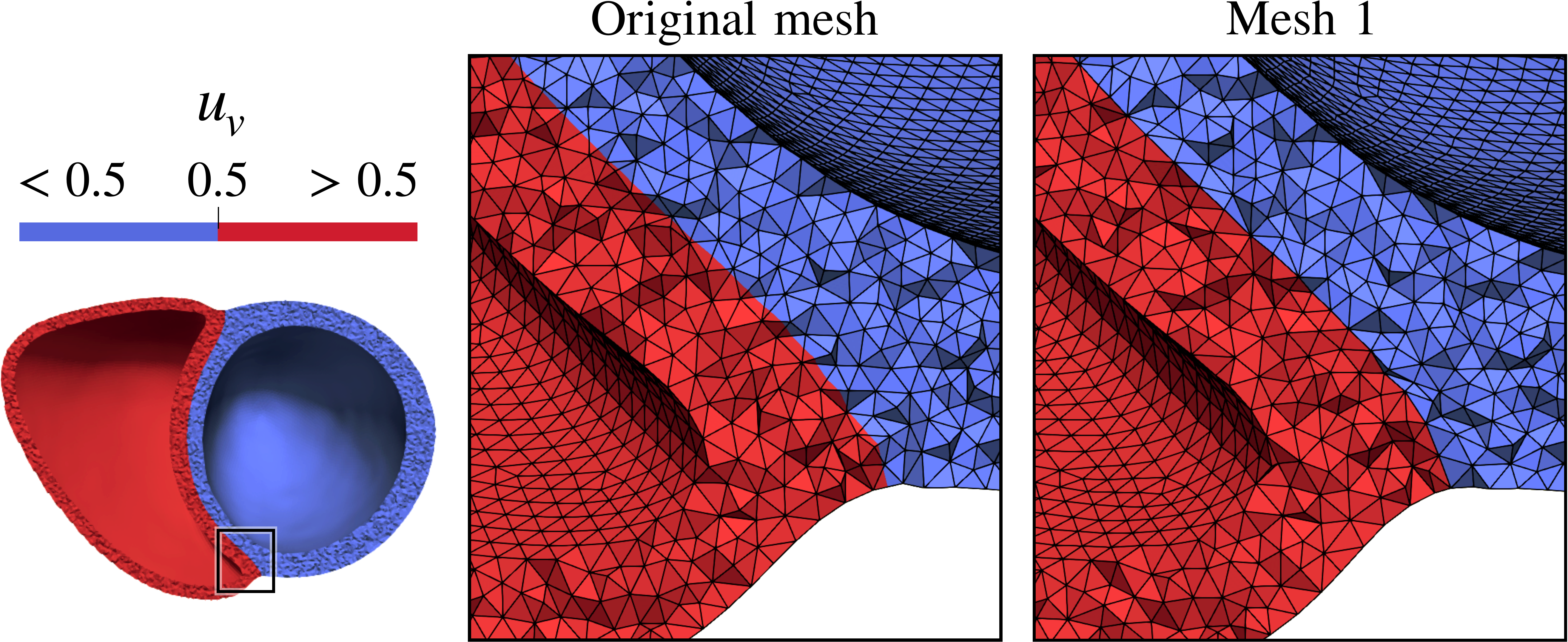}
	\caption{Close-up of the clipped mesh at the anterior interventricular junction before (\textit{left}) and after (\textit{right}) isovalue discretization at $u_v=0.5$.}
	\label{fig:mesh1}
\end{figure}

For being able to apply boundary conditions exactly at the boundary between the LV and the RV, we perform isovalue discretization at $u_v=0.5$, which yields \textit{mesh~1} (Fig.~\ref{fig:mesh1}). This means that the original tetrahedral mesh is remeshed, such that there are nodes directly on the boundary between the two ventricles.

\begin{figure}[H]
	\centering
	\includegraphics[width=\linewidth]{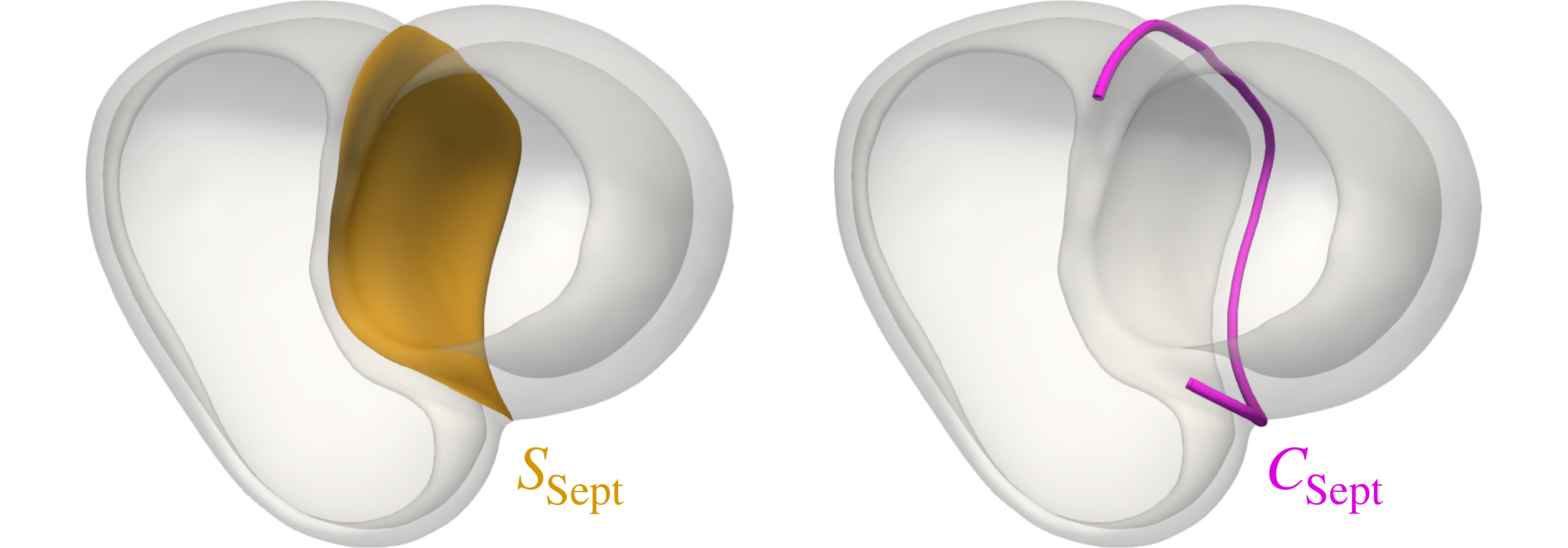}
	\caption{Septal surface (\textit{left}) and septal curve (\textit{right}).}
	\label{fig:septSurfaceCurve}
\end{figure}

\noindent From mesh~1, we extract all tetrahedron faces composing this boundary, which results in a \textit{septal surface} $S_\mathrm{Sept}$ (Fig.~\ref{fig:septSurfaceCurve},~left):
\begin{equation}
	S_\mathrm{Sept} = \bigl\{\mathbf{x} \in V \mid u_v(\mathbf{x}) = 0.5\bigr\}
\end{equation}
Similarly, we can extract a \textit{septal curve} $C_\mathrm{Sept}$ from the corresponding epicardial surface (Fig.~\ref{fig:septSurfaceCurve},~right):
\begin{equation}
	C_\mathrm{Sept} = \bigl\{\mathbf{x} \in S_\mathrm{Epi} \mid u_v(\mathbf{x}) = 0.5\bigr\}
\end{equation}

\subsubsection{Computation of transmural coordinate $m$}
\label{transmural}
\begin{figure}[H]
	\vspace{1mm}
	\centering
	\includegraphics[width=\linewidth]{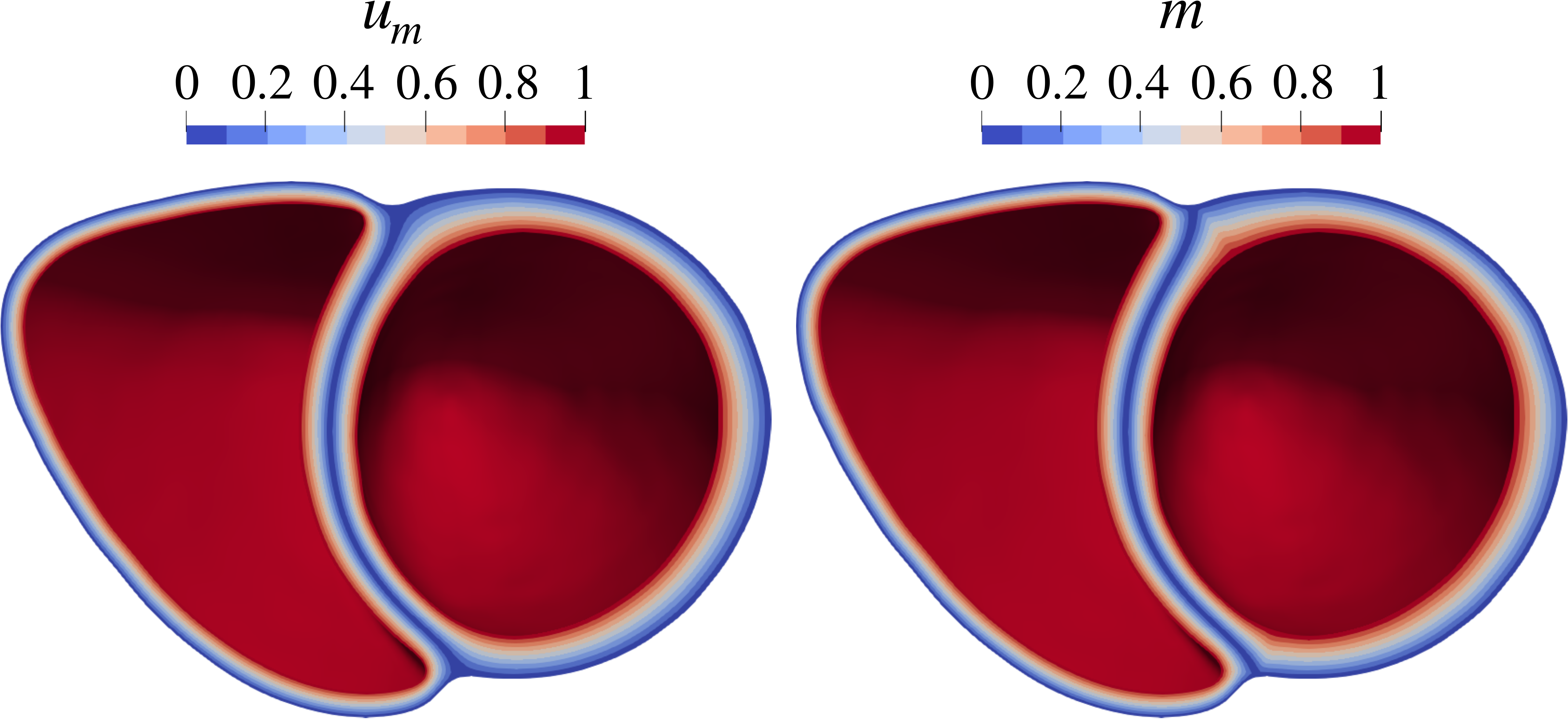}
	\caption{Computation of the transmural coordinate. \textit{Left:} Laplace solution. \textit{Right:} Final coordinate. The geometry was clipped for visualization.}
	\label{fig:transmural}
\end{figure}

To obtain the transmural coordinate, we first compute a Laplace solution that is 0 at the epicardial and the septal surface and 1 at the LV and RV endocardial surfaces (Fig.~\ref{fig:transmural}, left):
\begin{alignat}{2}
	\Delta u_m(V) = 0 \quad & \text{with}\quad & u_m(S_\mathrm{Epi}\cup S_\mathrm{Sept}) &= 0 \label{eq:transmuralLap}\\
	& \text{and}\quad & u_m(S_\mathrm{LV}\cup S_\mathrm{RV}) &= 1 \nonumber
\end{alignat}
Next, we compute trajectory distances $d_m$ along the gradient of this Laplace solution in both directions, i.e., starting from the epicardium and starting from the endocardium:
\begin{alignat}{2}
	\nabla d_{m,\mathrm{Epi}} \cdot \mathbf{t}_m &= 1 \quad\text{with}\quad &d_{m,\mathrm{Epi}}(S_\mathrm{Epi}\cup S_\mathrm{Sept}) = 0\\
	-\nabla d_{m,\mathrm{Endo}} \cdot \mathbf{t}_m &= 1 \quad\text{with}\quad &d_{m,\mathrm{Endo}}(S_\mathrm{LV}\cup S_\mathrm{RV}) = 0
\end{alignat}
\vspace{-\baselineskip}
\begin{equation}
	\text{where}\quad \mathbf{t}_m = \frac{\nabla u_m}{\|\nabla u_m\|} \label{eq:transmuralTangentField}
\end{equation}
The relative trajectory distance with respect to the epicardium is then defined as the \textit{transmural coordinate} $m$ (Fig.~\ref{fig:transmural}, right):
\begin{equation}
	m = \frac{d_{m,\mathrm{Epi}}}{d_{m,\mathrm{Epi}}+d_{m,\mathrm{Endo}}}
	\label{eq:transmuralCoord}
\end{equation}
Equations \eqref{eq:transmuralLap}-\eqref{eq:transmuralCoord} are solved on mesh 1 and the transmural coordinate is transferred back to the original tetrahedral mesh. To this end, we use linear interpolation, because the coordinates are spatially low-frequent and can well be approximated locally by a linear function.

\vspace{1mm}
\subsubsection{Extraction of heart axes and apex point}
\label{heartAxesApex}
\begin{figure}[t]
	\centering
	\includegraphics[width=\linewidth]{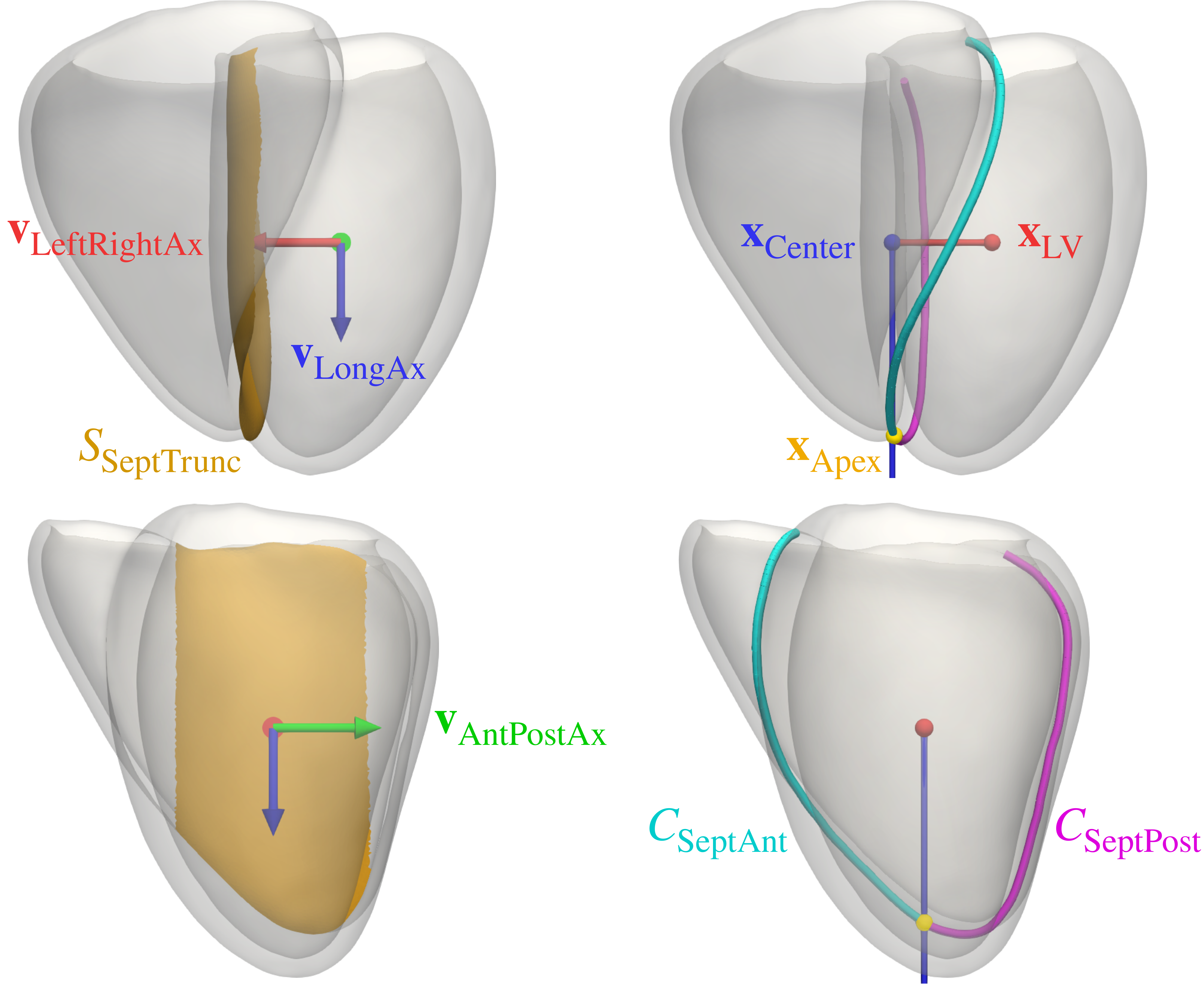}
	\caption{\textit{Left:} Truncated septal surface (dark yellow) and heart axes (blue, red and green arrows). \textit{Right:} Steps to locate the apex point: The LV centroid (red dot) is projected (red line) onto the plane defined by the left-right axis and the septal centroid, which yields a global center point (blue dot). A line in long axis direction (blue) is starting at this center point and the point of the septal curve closest to this line is identified as apex point (yellow dot). The apex point splits the septal curve into an anterior part (cyan) and a posterior part (magenta).}
	\label{fig:heartAxesApex}
\end{figure}

For the rotational and apicobasal coordinate, a consistent and robust definition of an epicardial apex point is essential. As this point will be used to define the apex for both ventricles, it should lie at the center between the two ventricles. Therefore, possible points are restricted to the septal curve in Fig.~\ref{fig:septSurfaceCurve}~(right). The most straightforward choice would be the point on this septal curve with the maximum distance to the basal surface. However, this definition would not be very robust, as the position along the septal curve would largely depend on its local shape and smoothness in the apex region. To yield an intuitive course of the rotational coordinate in the LV, the apex point should furthermore be centered with the LV in anterior-posterior direction. For this reason, we decided to take a more global approach that relies on the definition of orthogonal heart axes as depicted in the left half of Fig.~\ref{fig:heartAxesApex}.

The \textit{long axis} $\mathbf{v}_\mathrm{LongAx}$ is defined as the unit vector ``most orthogonal'' to the normals of the LV endocardial surface, as measured by the dot product with all triangle normals $\mathbf{n}_\mathrm{LV}$:
\begin{equation}
	\mathbf{v}_\mathrm{LongAx} = \underset{\mathbf{v}\in \mathbb{R}^3}{\operatorname{arg\,min}}\ \|\mathbf{v} \cdot \mathbf{n}_\mathrm{LV}\|_p
	\label{eq:longAxis}
\end{equation}
Here, $\|\cdot\|_p$ denotes the $p$-norm across all surface triangles. We chose $p=1.373$. % $p=(2-\log_2(1+\sqrt{2}))^{-1}$
For this value, the norm's unit circle lies at the center between those for $p=1$ and $p=2$. Similar values work as well.
Problem~\eqref{eq:longAxis} is solved using the Nelder-Mead algorithm.
To assure that the long axis is directed from base towards apex, its dot product with the vector pointing from the centroid of the basal surface to the centroid of the LV endocardial surface is evaluated and the long axis is flipped accordingly.

The definition of the left-right axis is based on fitting a plane to the septal surface in Fig.~\ref{fig:septSurfaceCurve}~(left). As the septal surface may become strongly curved near the interventricular junctions, particularly at the anterior side, a two-step process is used to only take into account the central part of the septal surface.\\
In the first step, principal component analysis is applied to the points on the entire septal surface. The third principal component represents the normal vector of the best-fitting plane and is defined as $\mathbf{v}_\mathrm{LR,Entire}$. Here, the vector pointing from the centroid of the LV endocardial surface to the centroid of the RV endocardial surface is used as reference to assure that this vector is directed from left to right. The distance in the direction of $\mathbf{v}_\mathrm{AP} = \mathbf{v}_\mathrm{LongAx} \times \mathbf{v}_\mathrm{LR,Entire}$ is then used to truncate the septal surface by $20\,\%$ and $10\,\%$ at the anterior and posterior side, respectively, which yields the \textit{truncated septal surface} $S_\mathrm{SeptTrunc}$:
\begin{align}
	S_\mathrm{SeptTrunc} = \bigl\{\mathbf{x} \in S_\mathrm{Sept} \mid\ 
	&\mathbf{x}\cdot\mathbf{v}_\mathrm{AP} > \operatorname{P}_{20}(\mathbf{x}\cdot\mathbf{v}_\mathrm{AP}) \text{ and}\\
	&\mathbf{x}\cdot\mathbf{v}_\mathrm{AP} < \operatorname{P}_{90}(\mathbf{x}\cdot\mathbf{v}_\mathrm{AP})\bigr\} \nonumber
\end{align}
Here, $\operatorname{P}_q$ denotes the $q$\textsuperscript{th} percentile.\\
In the second step, the final \textit{left-right axis} $\mathbf{v}_\mathrm{LeftRightAx}$ is obtained by computing the third principal component $\mathbf{v}_\mathrm{LR,Trunc}$ of points on the truncated septal surface and orthogonalizing it with respect to the long axis:
\begin{equation}
	\mathbf{v}_\mathrm{LeftRightAx} = \mathbf{v}_\mathrm{LR,Trunc} - (\mathbf{v}_\mathrm{LR,Trunc}\cdot\mathbf{v}_\mathrm{LongAx})\,\mathbf{v}_\mathrm{LongAx}
\end{equation}

The \textit{anterior-posterior axis} $\mathbf{v}_\mathrm{AntPostAx}$ is finally defined as:
\begin{equation}
	\mathbf{v}_\mathrm{AntPostAx} = \mathbf{v}_\mathrm{LongAx} \times \mathbf{v}_\mathrm{LeftRightAx}
\end{equation}

\newpage
The three heart axes are then used to find the apex point. This is illustrated in the right half of Fig.~\ref{fig:heartAxesApex}. First, a global \textit{center point} $\mathbf{x}_\mathrm{Center}$ is obtained by projecting the centroid $\mathbf{x}_\mathrm{LV}$ of the LV endocardial surface onto the plane perpendicular to the left-right axis that passes through the centroid $\mathbf{x}_\mathrm{SeptTrunc}$ of the truncated septal surface (red line):
\begin{equation}
	\mathbf{x}_\mathrm{Center} = \mathbf{x}_\mathrm{LV} + ((\mathbf{x}_\mathrm{SeptTrunc}-\mathbf{x}_\mathrm{LV}) \cdot \mathbf{v}_\mathrm{LeftRightAx})\,\mathbf{v}_\mathrm{LeftRightAx}
\end{equation}
The \textit{apex point} $\mathbf{x}_\mathrm{Apex}$ is then located as the point of the septal curve with the smallest distance to the line in long axis direction starting at this center point (blue line):
\begin{gather}
	\mathbf{x}_\mathrm{Apex} = \underset{\mathbf{x} \in C_\mathrm{Sept}}{\operatorname{arg\,min}}\  \|\mathbf{x}+r_\mathrm{LongAx}\,\mathbf{v}_\mathrm{LongAx}-\mathbf{x}_\mathrm{Center}\|\\
	\text{with}\quad r_\mathrm{LongAx} = (\mathbf{x}_\mathrm{Center}-\mathbf{x})\cdot\mathbf{v}_\mathrm{LongAx} > 0\nonumber
\end{gather}
This apex point is used to split the septal curve into an \textit{anterior septal curve} $C_\mathrm{SeptAnt}$ and a \textit{posterior septal curve} $C_\mathrm{SeptPost}$.

\subsubsection{Second remeshing; extraction of ridge surfaces}
\label{mesh2}
\begin{figure}[H]
	\centering
	\includegraphics[width=\linewidth]{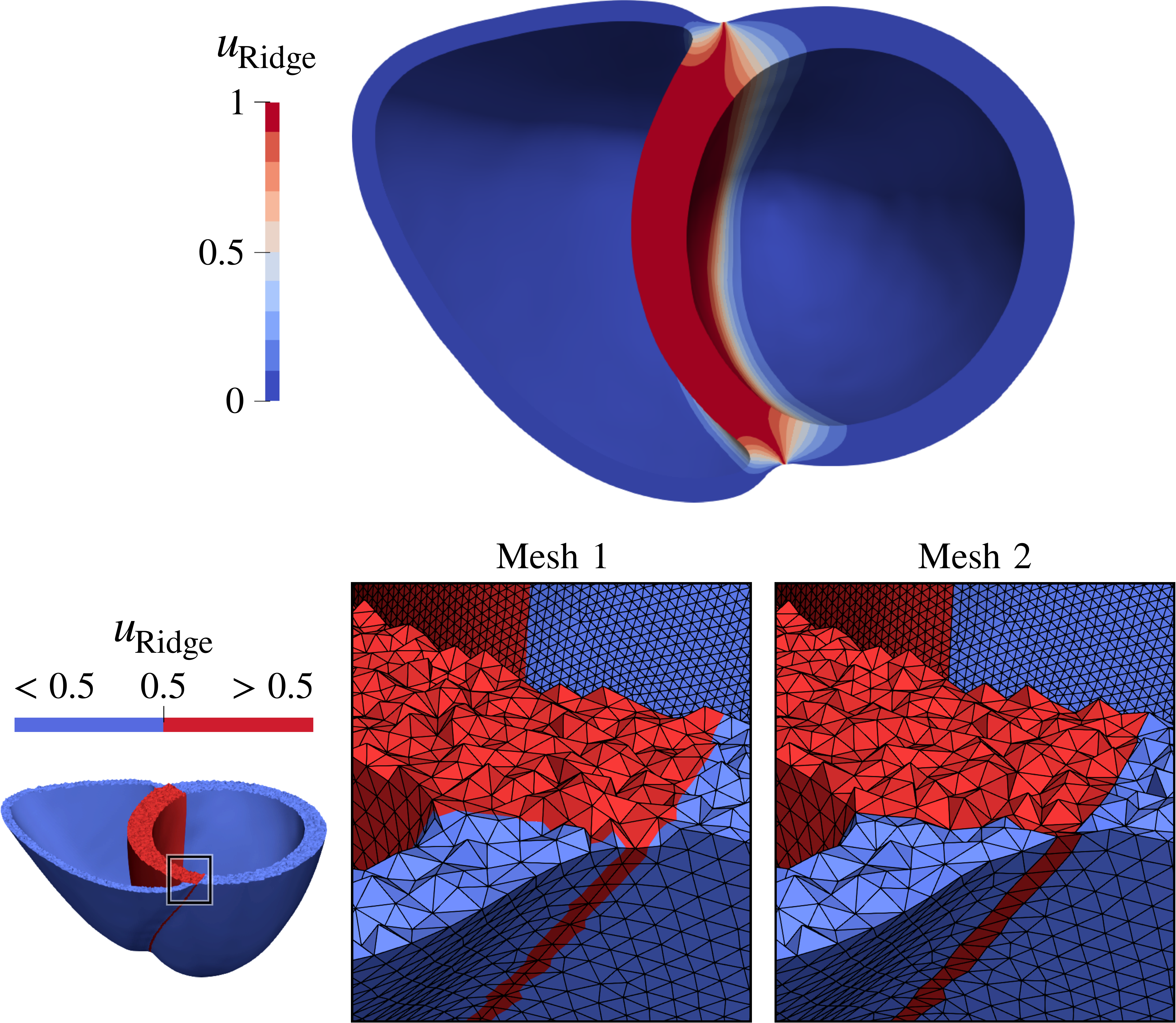}
	\caption{\textit{Upper half:} ``Ridge'' Laplace solution. The geometry was clipped for visualization. \textit{Lower half:} Close-up of the mesh at the anterior septal junction before (\textit{left}) and after (\textit{right}) isovalue discretization at $u_\mathrm{Ridge}=0.5$.}
	\label{fig:mesh2}
\end{figure}

Computing a rotational coordinate by solving a PDE requires to define at least two surfaces for assigning boundary conditions. For consistency, these surfaces should be based on anatomical landmarks that can be identified reliably on different geometries. As we furthermore aim for a rotational coordinate that is symmetric in both ventricles and that allows to distinguish between the septum and the free walls, the anterior and posterior junctions between the septum and both free walls are a natural choice for such landmarks. To obtain boundary surfaces representing these two junctions, we first compute a Laplace solution that is 0 on the epicardial surface and 1 on the septal surface (see upper half of Fig.~\ref{fig:mesh2}):
\begin{alignat}{2}
	\Delta u_\mathrm{Ridge}(V) = 0 \quad & \text{with}\quad & u_\mathrm{Ridge}(S_\mathrm{Epi} \setminus S_\mathrm{Sept}) &= 0
	\label{eq:ridgeLaplace}\\
	& \text{and}\quad & u_\mathrm{Ridge}(S_\mathrm{Sept}) &= 1 \nonumber
\end{alignat}
\noindent Then we perform isovalue discretization at $u_\mathrm{Ridge} = 0.5$, which yields \textit{mesh 2} (lower half of Fig.~\ref{fig:mesh2}). Note that in the boundary conditions of \eqref{eq:ridgeLaplace}, the epicardial points of the septal surface are excluded from the epicardial surface to obtain disjoint mesh regions (no common nodes) for the left and right free walls after remeshing.

\begin{figure}[H]
	\centering
	\includegraphics[width=\linewidth]{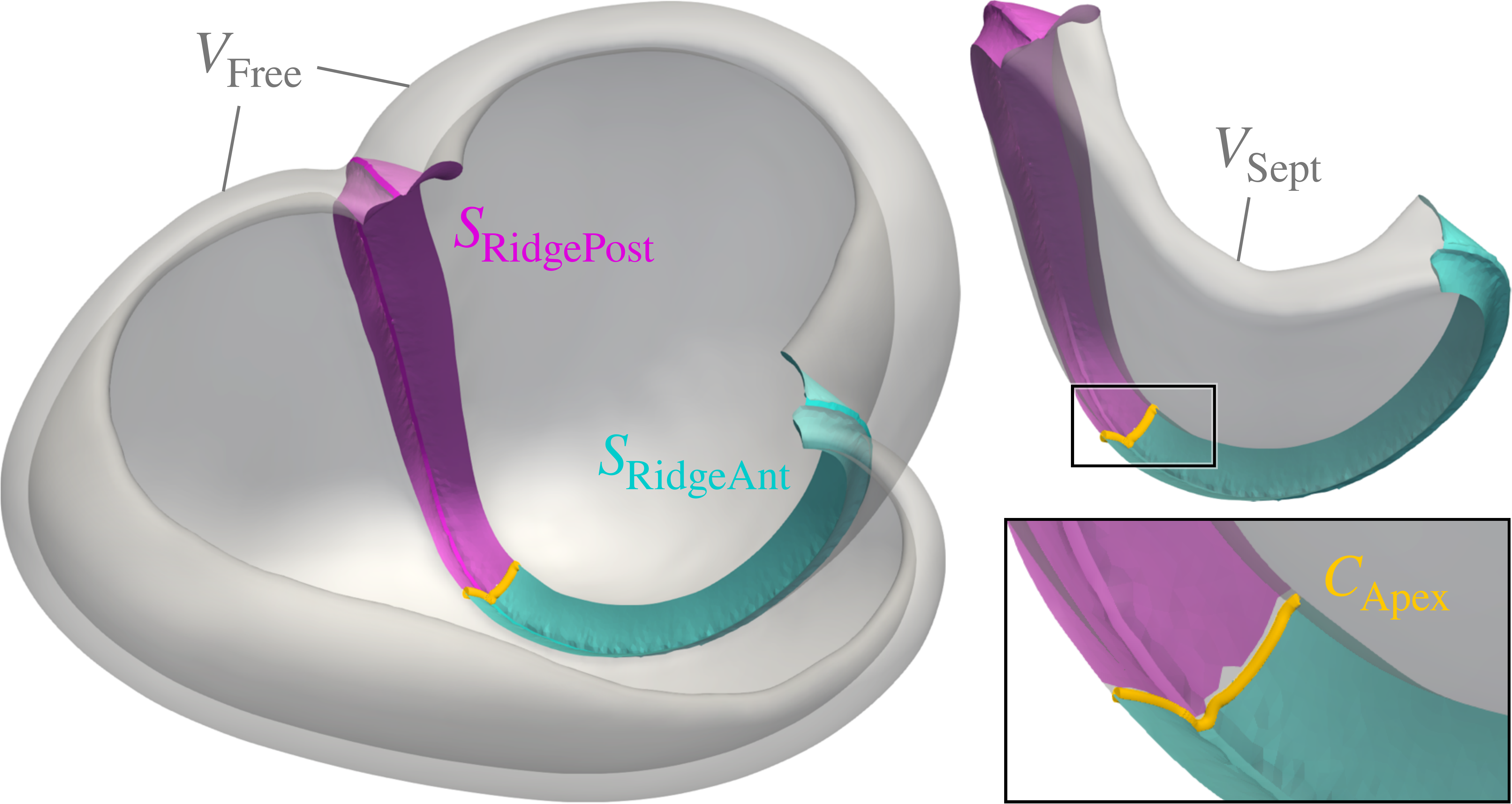}
	\caption{Anterior and posterior ridge surfaces and apex curve. The apex curve runs from the epicardium to the LV and RV endocardium.}
	\label{fig:ridgeSurfaces}
\end{figure}

From mesh~2, we extract the volume $V_\mathrm{Free}$ covering both free walls, the volume $V_\mathrm{Sept}$ covering the septal wall, and a \textit{ridge surface}~$S_\mathrm{Ridge}$:
\begin{align}
	V_\mathrm{Free} &= \bigl\{\mathbf{x} \in V \mid u_\mathrm{Ridge}(\mathbf{x}) \leq 0.5\bigr\}\\
	V_\mathrm{Sept} &= \bigl\{\mathbf{x} \in V \mid u_\mathrm{Ridge}(\mathbf{x}) \geq 0.5\bigr\}\\
	S_\mathrm{Ridge} &= \bigl\{\mathbf{x} \in V \mid u_\mathrm{Ridge}(\mathbf{x}) = 0.5\bigr\}
\end{align}
The shortest Euclidean distance to the anterior and posterior septal curves is used to split the ridge surface into an \textit{anterior ridge surface} $S_\mathrm{RidgeAnt}$ and a \textit{posterior ridge surface} $S_\mathrm{RidgePost}$ in a nearest neighbor manner. Furthermore, a transmural \textit{apex curve} $C_\mathrm{Apex}$ is obtained between these two surfaces (Fig.~\ref{fig:ridgeSurfaces}):
\begin{align}
	S_\mathrm{RidgeAnt} &= \bigl\{\mathbf{x} \in S_\mathrm{Ridge} \mid r_\mathrm{Ant}(\mathbf{x}) < r_\mathrm{Post}(\mathbf{x})\bigr\}\\
	S_\mathrm{RidgePost} &= \bigl\{\mathbf{x} \in S_\mathrm{Ridge} \mid r_\mathrm{Ant}(\mathbf{x}) > r_\mathrm{Post}(\mathbf{x})\bigr\}\\
	C_\mathrm{Apex} &= \bigl\{\mathbf{x} \in S_\mathrm{Ridge} \mid r_\mathrm{Ant}(\mathbf{x}) = r_\mathrm{Post}(\mathbf{x})\bigr\}
\end{align}
\vspace{-1.5\baselineskip}
\begin{equation*}
	\text{with}\quad r_\mathrm{Ant}(\mathbf{x}) = \min_{\mathbf{y}\in C_\mathrm{SeptAnt}}\|\mathbf{x}-\mathbf{y}\|, \quad r_\mathrm{Post}(\mathbf{x}) = \min_{\mathbf{y}\in C_\mathrm{SeptPost}}\|\mathbf{x}-\mathbf{y}\|
\end{equation*}
As there are no nodes that exactly fulfill $r_\mathrm{Ant}(\mathbf{x}) = r_\mathrm{Post}(\mathbf{x})$, the closest nodes on the anterior ridge surface define the apex curve in the discrete mesh.

\subsubsection{Computation of rotational coordinate $r$}
\label{rotational}
\begin{figure}[H]
	\centering
	\includegraphics[width=\linewidth]{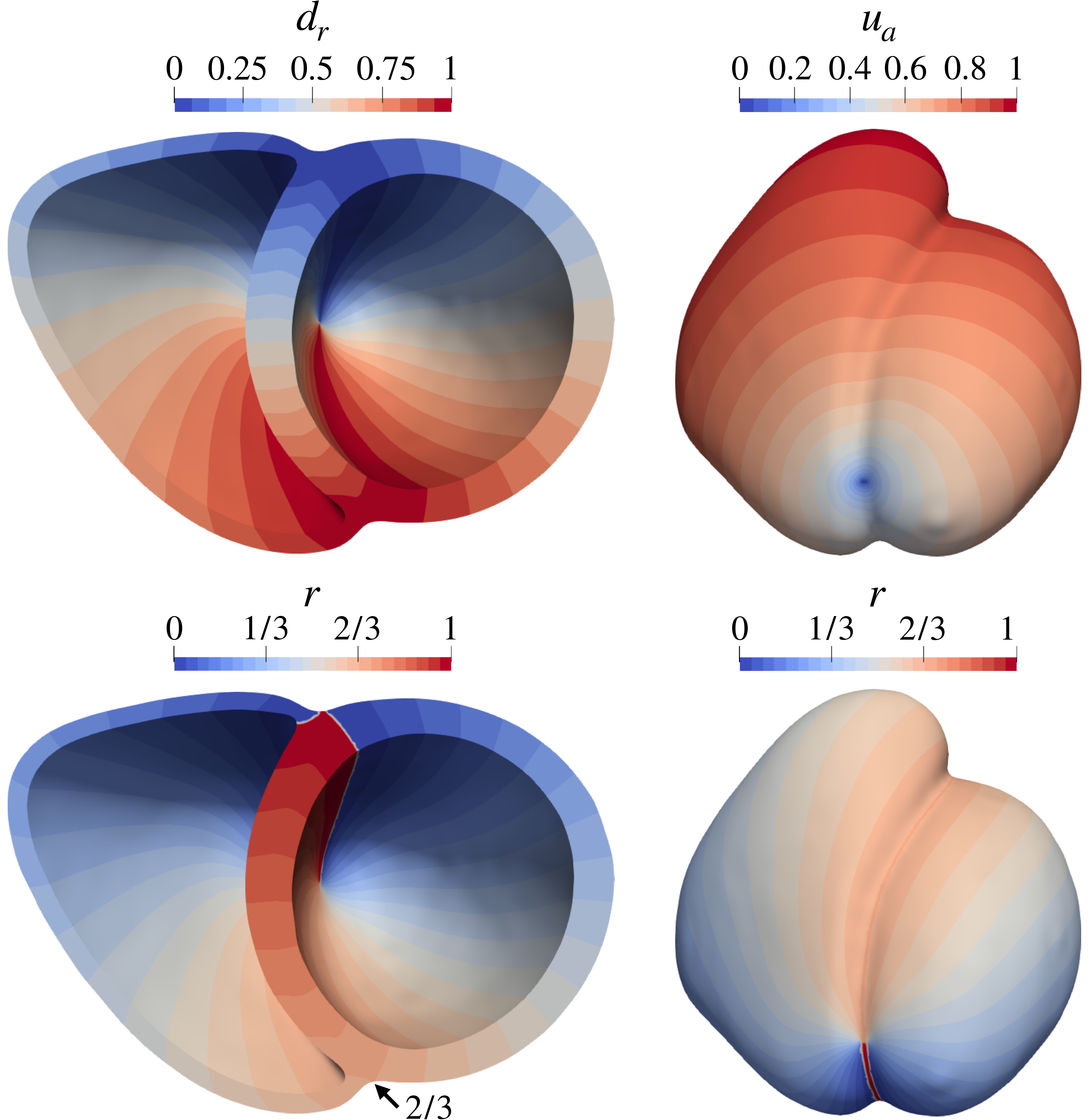}
	\caption{Computation of the rotational coordinate. \textit{Top-left:} Relative trajectory distance. \textit{Top-right:} Apicobasal Laplace solution. \textit{Bottom:} Final coordinate. The geometry shown on the left was clipped for visualization.}
	\label{fig:rotational}
\end{figure}

The relative trajectory distance between the posterior and anterior ridge surfaces is used to define the rotational coordinate (Fig.~\ref{fig:rotational}, top-left). It is computed separately within the free walls and the septum:
\begin{align}
	d_r(V_\mathrm{Free}) &= \frac{d_{r,\mathrm{Post}}(V_\mathrm{Free})}{d_{r,\mathrm{Post}}(V_\mathrm{Free})+d_{r,\mathrm{Ant}}(V_\mathrm{Free})} \label{eq:rotTrajectDistFree}\\
	d_r(V_\mathrm{Sept}) &= \frac{d_{r,\mathrm{Post}}(V_\mathrm{Sept})}{d_{r,\mathrm{Post}}(V_\mathrm{Sept})+d_{r,\mathrm{Ant}}(V_\mathrm{Sept})}
\end{align}
where $d_{r,\mathrm{Post}}$ and $d_{r,\mathrm{Ant}}$ are given by:
\vspace{-1.5\baselineskip}
\begin{alignat}{2}
	\nabla d_{r,\mathrm{Post}}(V_\mathrm{Free}) \cdot \mathbf{t}_r(V_\mathrm{Free}) &= 1 \hspace{0.5em}\text{with}\hspace{0.5em} &d_{r,\mathrm{Post}}(S_\mathrm{RidgePost}) = 0\\
	-\nabla d_{r,\mathrm{Ant}}(V_\mathrm{Free}) \cdot \mathbf{t}_r(V_\mathrm{Free}) &= 1 \hspace{0.5em}\text{with}\hspace{0.5em} &d_{r,\mathrm{Ant}}(S_\mathrm{RidgeAnt}) = 0\\
	\nabla d_{r,\mathrm{Post}}(V_\mathrm{Sept}) \cdot \mathbf{t}_r(V_\mathrm{Sept}) &= 1 \hspace{0.5em}\text{with}\hspace{0.5em} &d_{r,\mathrm{Post}}(S_\mathrm{RidgePost}) = 0\\
	-\nabla d_{r,\mathrm{Ant}}(V_\mathrm{Sept}) \cdot \mathbf{t}_r(V_\mathrm{Sept}) &= 1 \hspace{0.5em}\text{with}\hspace{0.5em} &d_{r,\mathrm{Ant}}(S_\mathrm{RidgeAnt}) = 0
\end{alignat}
Here, the tangent field $\mathbf{t}_r$ is not based on the gradient of a rotational Laplace solution but on the cross product of the gradients of the transmural coordinate $m'$ and an apicobasal Laplace solution $u_a$:
\begin{equation}
	\mathbf{t}_r = \frac{\nabla m'}{\|\nabla m'\|} \times \frac{\nabla u_a}{\|\nabla u_a\|}
	\label{eq:rotTangentField}
\end{equation}
The transmural coordinate is inverted in the RV to get coherent gradients in the septum and opposite directions of rotation in both ventricles:
\begin{equation}
	m'(\mathbf{x}) =
	\begin{cases}
		m(\mathbf{x}), & v(\mathbf{x})=0\\
		-m(\mathbf{x}), & v(\mathbf{x})=1\\
	\end{cases}
\end{equation}
As \eqref{eq:rotTrajectDistFree}-\eqref{eq:rotTangentField} are computed on mesh~2, linear interpolation is used to transfer $m'$ from mesh~1 to mesh~2.
The apicobasal Laplace solution (Fig.~\ref{fig:rotational}, top-right) is computed directly on mesh 2 and is 0 at the apex curve and 1 at the basal surface:
\begin{equation}
	\Delta u_a(V) = 0 \hspace{0.9em}\text{with}\hspace{0.9em} u_a(C_\mathrm{Apex})=0 \hspace{0.9em}\text{and}\hspace{0.9em} u_a(S_\mathrm{Base})=1
\end{equation}
The choice of the tangent field in \eqref{eq:rotTangentField} has two advantages over using a rotational Laplace solution. First, it does not lead to distortions of the resulting rotational coordinate near the base due to Neumann boundary conditions that would have to be imposed on the rotational Laplace solution. Second, the gradient direction of the apicobasal Laplace solution approximates the gradient direction of the final apicobasal coordinate and using the cross product between the transmural and apicobasal directions increases the linear independence of the rotational coordinate from these two coordinate directions.

The final \textit{rotational coordinate} $r$ (Fig.~\ref{fig:rotational}, bottom) is obtained by flipping, scaling and shifting the relative trajectory distances:
\begin{align}
	r(V_\mathrm{Free}) &= \tfrac{2}{3}\ d_r(V_\mathrm{Free})\\
	r(V_\mathrm{Sept}) &= \tfrac{2}{3} + \tfrac{1}{3}\ \bigl(1-d_r(V_\mathrm{Sept})\bigr)
\end{align}
Based on average geometrical proportions and in accordance with the ratio of two septal and four free wall segments in the AHA scheme \citep{cerqueira01}, the scaling factors were chosen such that the septum covers one third and the free walls two thirds of the total range $[0,1]$.
The rotational coordinate starts with 0 at the posterior septal junction, increases across the free walls up to a value of 2/3 at the anterior septal junction and then traverses the septum until it reaches the posterior septal junction once again, with a value of 1. The discontinuity at the posterior junction can be avoided by transforming the rotational coordinate into two continuous coordinates -- a \textit{rotational sine coordinate}~$r_\mathrm{sin}$ and a \textit{rotational cosine coordinate}~$r_\mathrm{cos}$:
\begin{align}
	r_\mathrm{sin} &= \sin(2\pi r)\label{eq:rtSin}\\
	r_\mathrm{cos} &= \cos(2\pi r)\label{eq:rtCos}
\end{align}
This trick is used for linear interpolation back onto the original mesh, where the following inverse transform is applied:
\begin{equation}
	r = \begin{cases}
		\operatorname{atan2}(r_\mathrm{sin}, r_\mathrm{cos})/(2\pi), & r_\mathrm{sin} \geq 0\\
		\operatorname{atan2}(r_\mathrm{sin}, r_\mathrm{cos})/(2\pi)+1, & r_\mathrm{sin} < 0
	\end{cases}
\end{equation}

\subsubsection{Computation of apicobasal coordinate $a$}
\label{apicobasal}
\begin{figure}[H]
	\centering
	\includegraphics[width=\linewidth]{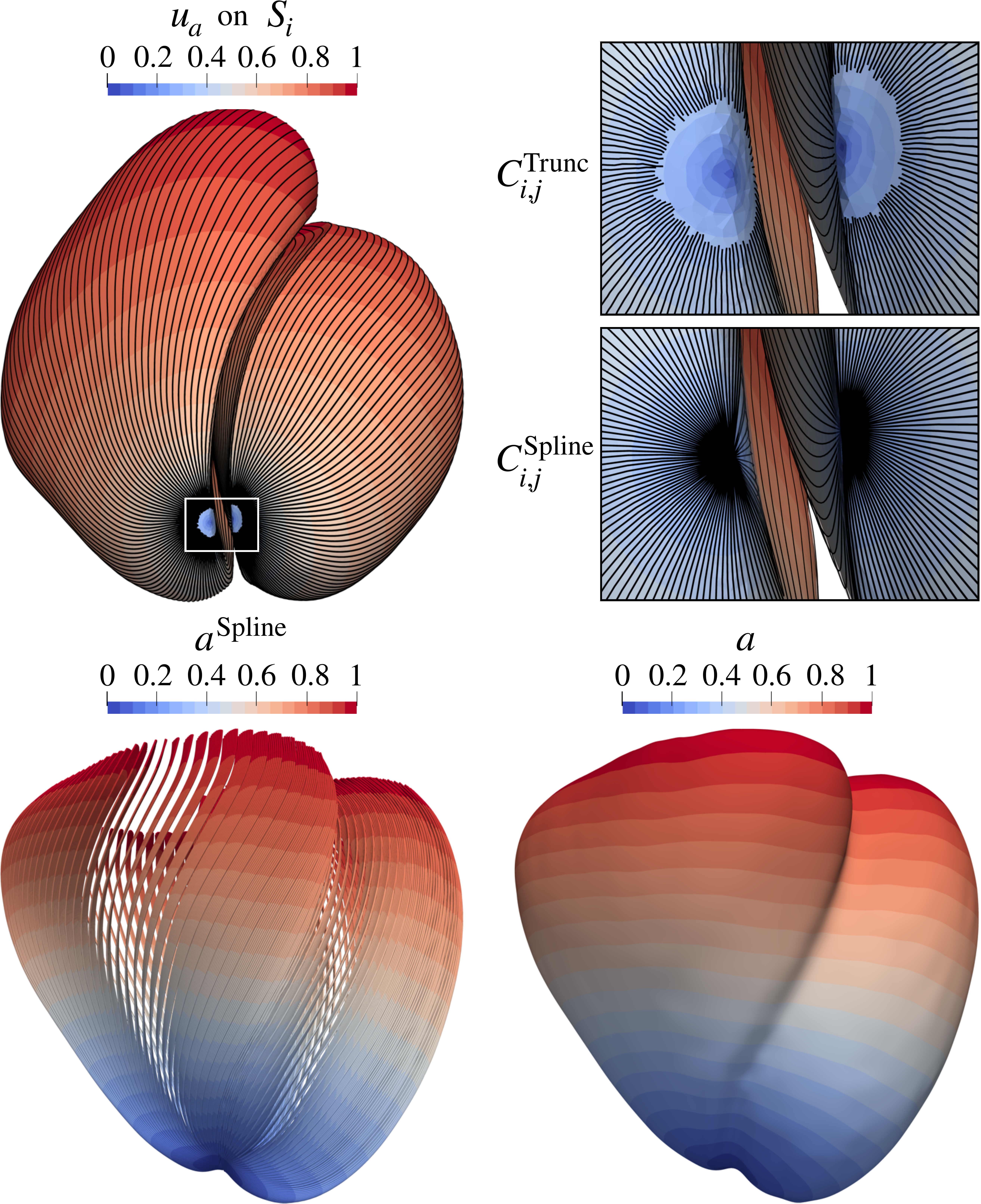}
	\vspace{-1mm}
	\caption{Computation of the apicobasal coordinate. \textit{Upper half:} Apicobasal Laplace solution on the LV and RV isosurfaces for one out of 10 transmural values and corresponding isocurves for all 96 rotational values (black lines) after truncation and spline fitting. \textit{Lower half:} Normalized distance on all 2$\cdot$10$\cdot$96 spline curves (\textit{left}) and final coordinate on the original mesh (\textit{right}).}
	\label{fig:apicobasal}
\end{figure}

Although trajectory distances between two boundary surfaces are used to define the transmural and the rotational coordinate, this approach is not well suited for the apicobasal coordinate. The reason is that in this case, we are looking for a normalized distance between the two-dimensional basal surface and the one-dimensional apex curve. Due to the different dimensionality of boundaries, trajectories starting at different points on the basal surface may end at the same point on the apex curve, which leads to contradicting values of the trajectory distance. Therefore, a different approach is used: Apicobasal curves are obtained by extracting isocontours at discrete values of the transmural and the rotational coordinate and the normalized distance along these curves is determined.\\
We start by extracting isosurfaces of the transmural coordinate from mesh 1. The following 10 isovalues are used to obtain an equidistant sampling:
\begin{equation}
	m \in \left\{\tfrac{1}{20}, \tfrac{3}{20}, \tfrac{5}{20}, \dots, \tfrac{19}{20}\right\}
\end{equation}
This results in 20 disjoint isosurfaces $S_i$ ($i=1,2,\dots,20$). There are twice as many isosurfaces as isovalues, because one surface per ventricle is extracted for each isovalue.\\
Next, isocurves of the rotational coordinate are extracted from each of these isosurfaces. To this end, the rotational sine and cosine coordinates are linearly interpolated from mesh 2 to the isosurfaces. The following 96 isovalues are chosen to yield a sufficiently fine sampling that captures the septal junctions at $r=2/3$ and $r=1$:
\begin{equation}
	r \in \left\{\tfrac{1}{96}, \tfrac{2}{96}, \tfrac{3}{96}, \dots, \tfrac{96}{96}\right\}
\end{equation}
This results in 1920 isocurves $C_{i,j}$ ($i$ as in $S_i$, $j=1,2,\dots,96$).\\
However, all these curves are connected at the apex region. As disconnected curves with a well-defined apical start point are required to determine normalized distances along the curves, a few more intermediate steps are necessary, which are illustrated in the upper half of Fig.~\ref{fig:apicobasal}.\\
To disconnect the isocurves at the apex, the apicobasal Laplace solution is also interpolated onto the isosurfaces $S_i$ and one individual apex point $\mathbf{x}_i$ is determined for each $S_i$ by finding the minimum of the Laplace solution. Then, the curves are truncated by excluding points within a radius $\varepsilon$ to the respective apex point:
\begin{equation}
	C_{i,j}^\mathrm{Trunc} = \bigl\{\mathbf{x} \in C_{i,j} \mid \|\mathbf{x}-\mathbf{x}_i\| > \varepsilon\bigr\}
	\hspace{2mm}\text{with}\hspace{2mm} \mathbf{x}_i = \underset{\mathbf{x} \in S_i}{\operatorname{arg\,min}}\ u_a(\mathbf{x})
\end{equation}
An $\varepsilon$ of three times the mean edge length of the original mesh was found to be sufficient to ensure disjoint curves.\\
To obtain smooth curves with well-defined apical start points, the corresponding $\mathbf{x}_{i}$ is re-added to each $C_{i,j}^\mathrm{Trunc}$ and a cubic smoothing spline fit~\citep{Reinsch1967} is used to resample each curve at 100 equidistant nodes along the curve. This yields the \textit{spline curves}~$C_{i,j}^\mathrm{Spline}$. The extent of smoothing is determined such that the root-mean-square deviation (RMSD) from the original points equals $0.5\,\%$ of the apicobasal distance to strike a balance between smoothness and the original course. To enforce that each spline curve passes through the respective~$\mathbf{x}_{i}$, a 100-fold weight is used for this point.
The normalized distance $a^\mathrm{Spline}$ along each spline curve is then computed as the relative cumulative sum of Euclidean distances between neighboring nodes on this curve, starting at~$\mathbf{x}_i$. The result can be seen at the bottom-left of Fig.~\ref{fig:apicobasal}.\\
Laplacian extrapolation is used to obtain the \textit{apicobasal coordinate}~$a$ on mesh 1 from $a^\mathrm{Spline}$:
\begin{align}
	\mathbf{a} &= \underset{\mathbf{a}}{\operatorname{arg\,min}}\left(\|\mathbf{R}\mathbf{a}-\mathbf{a^\mathrm{Spline}}\|^2 + \lambda\,\|\mathbf{L}\mathbf{a}\|^2 + \eta\,\|\mathbf{E}\mathbf{a}-\mathbf{1}\|^2\right) \label{eq:lapExtrap}\\
	&= (\mathbf{R}^\mathsf{T}\mathbf{R} + \lambda\,\mathbf{L}^\mathsf{T}\mathbf{L} + \eta\,\mathbf{E}^\mathsf{T}\mathbf{E})^{-1} (\mathbf{R}^\mathsf{T}\mathbf{a^\mathrm{Spline}} + \eta\,\mathbf{E}^\mathsf{T}\mathbf{1}) \nonumber
\end{align}
Here, the vector $\mathbf{a^\mathrm{Spline}}$ contains $a^\mathrm{Spline}$ at all nodes of the spline curves and $\mathbf{a}$ contains $a$ at all nodes of the volume mesh. $\mathbf{R}$ is a matrix that linearly interpolates from the nodes of the volume mesh onto the nodes of the spline curves and $\mathbf{L}$ is the Laplacian operator of the volume mesh. The smoothing parameter~$\lambda$ is determined using the secant method, such that the RMSD between $\mathbf{R}\mathbf{a}$ and $\mathbf{a}^\mathrm{Spline}$ equals $0.25\,\%$. The last term in \eqref{eq:lapExtrap} forces the extrapolated values to 1 at the base. $\mathbf{E}$ extracts the values at the basal surface of the volume mesh and $\eta$ is chosen to yield an equal weighting with the first term:
\begin{equation}
	\eta = \left(\tfrac{\mathrm{number\ of\ nodes\ on\ the\ spline\ curves}}{\mathrm{number\ of\ nodes\ on\ the\ basal\ surface}}\right)^2
\end{equation}
Finally, linear interpolation is used to transfer the apicobasal coordinate from mesh 1 to the original mesh. The result is depicted at the bottom-right of Fig.~\ref{fig:apicobasal}.

\subsection{Transferring data using Cobiveco}
\label{transferringData}
To transfer scalar data using Cobiveco, we construct a transfer matrix $\mathbf{M}_{B\leftarrow A}$ that maps from the nodes of a source mesh $A$ to the nodes of a target mesh $B$ (or a target point cloud). The principal transfer procedure is similar to the one described in~\cite{Bayer-2018-ID11708} -- with three advancements:
\begin{itemize}
	\item The discontinuity of the rotational coordinate is completely avoided by transforming it into the continuous sine and cosine coordinates using \eqref{eq:rtSin} and \eqref{eq:rtCos}.
	\item Mesh dependent instead of fixed scaling factors are used to yield scaled coordinates that show a comparable change per unit length in Euclidean space. To this end, the maximum coordinate difference of $m$, $r_\mathrm{sin}$, $r_\mathrm{cos}$ and $a$ between any two nodes of each tetrahedron is computed for $A$. The coordinates in both $A$ and $B$ are then divided by the median value of the respective maximum differences. As this is not possible for the binary coordinate, $v$ is instead multiplied by the bounding box diagonal and divided by the mean edge length of $A$.
	\item The rotational coordinates are additionally scaled as a function of the apicobasal coordinate. This is important to assure a well-defined mapping at the rotational singularities and to account for the decreasing circumference of the ventricles towards the apex. As the rotational coordinate becomes undefined at the apex curve (see Fig.~\ref{fig:ridgeSurfaces}~and~\ref{fig:rotational}), its weighting should become zero for $a=0$. As the circumference of the ventricles is roughly proportional to the square root of the apicobasal coordinate, $\sqrt{a}$ is chosen as scaling function for $r_\mathrm{sin}$ and $r_\mathrm{cos}$.
\end{itemize}
The data transfer functionality is also implemented in MATLAB. The user can choose between a transfer using linear or nearest-neighbor interpolation. For linear interpolation, the computation of $\mathbf{M}_{B\leftarrow A}$ consists of five steps:
\begin{enumerate}
	\item The coordinates are scaled as described above.
	\item For each node in $B$, the tetrahedron centroid in $A$ with the closest ventricular coordinates is found. This is done using a nearest neighbor search with a $k$-d tree and an Euclidean distance metric.
	\item For each centroid found in step 2, all centroids within a predefined search radius are found. This is done separately for the left and right ventricle using a range search with a $k$-d tree and an Euclidean distance metric. A search radius of two mean edge lengths of $A$ was found to be sufficient and used in this work.
	\item For each node in $B$, we iterate over the tetrahedrons corresponding to the respective centroids found in step 3. For each tetrahedron, we compute the barycentric coordinates that reproduce the ventricular coordinates of $B$. The tetrahedron with the smallest maximum absolute deviation of barycentric coordinates from $0.5$ is identified as tetrahedron to be used for interpolation.
	\item The barycentric coordinates and the node indices of tetrahedrons identified in step 4 are used to assemble $\mathbf{M}_{B\leftarrow A}$.
\end{enumerate}
For nearest-neighbor interpolation, steps 2--4 are replaced by directly finding nodes instead of tetrahedron centroids and $\mathbf{M}_{B\leftarrow A}$ is made up of ones instead of barycentric coordinates (step 5).

%%%%%%%%%%%%%%%%%%%%%%%%%%%%%%%%%%%%%%%%%%%%%%%%%%
\section{Evaluation}
\label{evaluation}

\subsection{Test geometries}
\label{testGeometries}

Two sets of biventricular geometries were used to evaluate Cobiveco: Geometries created using a statistical shape model (SSM) and imaged patient geometries.

\subsubsection{Statistical shape model}
The mean shape of the SSM from~\cite{Bai-2015-ID12225,deMarvao-2014-ID15002} was used as a representative geometry to evaluate transfer errors and 1000 quasi-random instances of this model were used to assess the computational robustness of Cobiveco. This SSM was created from more than 1000 magnetic resonance images. Originally, it consists of disconnected surfaces of the LV endo- and epicardium and the RV blood pool. To derive a model that can be used to compute coordinates, we extruded the RV blood pool by 3\,mm to obtain an RV epicardial surface and merged all surfaces to form one closed surface of the biventricular myocardium. This surface was tetrahedralized and the 100 principal components and variances were interpolated to the nodes of the volume mesh. The adapted model is publicly available \citep{Schuler2021shapemodel}.
Mesh statistics for the mean shape depicted in Fig.~\ref{fig:boundarySurfaces} can be found in the first row of Table~\ref{tab:meshStatistics}.
The 1000 quasi-random instances were created by drawing the weights of the 100 shape modes from a uniform distribution within bounds of $\pm 3$ standard deviations.

\subsubsection{Patient geometries}
36 patient geometries were used for a comparison of Cobiveco and UVC under realistic conditions. These geometries were acquired as part of validation studies~\citep{revishvili15a,chmelevsky2018} for electrocardiographic imaging (ECGI), which adhered to the Declaration of Helsinki and were approved by the Institutional Review Board of Almazov National Medical Research Center in Saint Petersburg, Russia. Written informed consent was obtained from each patient. Cardiac computed tomography (CT) images were obtained from patients with implanted pacemakers and segmented in a semi-automatic manner with the software of the Amycard 01C EP system (EP Solutions SA, Yverdon-les-Bains, Switzerland). As this system uses relatively coarse triangle meshes suitable for ECGI (edge lengths of 5 to 10\,mm), they were first remeshed with Instant Meshes~\citep{Jakob-2015-ID12648} and then tetrahedralized with Gmsh~\citep{Geuzaine-2009-ID12650}. Some geometries included large parts of the aorta and the pulmonary artery. To yield consistent inputs for the computation of coordinates, we clipped all meshes at the base (where the LV outflow tract intersects the septal plane) and removed the bridge at the base of the RV. All 36 geometries are shown in Fig.~\ref{fig:patient_cobiveco} and mesh statistics are given in Table~\ref{tab:meshStatistics}.

\subsection{Comparison with UVC}
\label{comparisonWithUVC}
For a comparison of Cobiveco with UVC, we also computed UVC coordinates for the mean shape of the SSM and all patient geometries. To this end, we reimplemented the UVC method in MATLAB according to the description in~\cite{Bayer-2018-ID11708}. This implementation is also accessible at \url{https://github.com/KIT-IBT/Cobiveco}. The UVC method was provided with the most comparable inputs: The epicardial apex point identified by Cobiveco was used as ``user-defined'' apex point and the part of the RV endocardial surface with $r\in[2/3,1]$ was used as RV septal surface.
To facilitate a direct comparison between Cobiveco and UVC, the original UVC coordinates $(\nu, \rho, \phi, \mathfrak{z})$ were transformed into coordinates $(v', m', r', a')$ that cover the same ranges as the corresponding Cobiveco coordinates (see Fig.~\ref{fig:concept}):
\begin{align}
	v' &= \tfrac{1}{2}+\tfrac{1}{2}\nu\\
	m' &= 1-\rho\\
	r' &= 
	\begin{cases}
		\frac{2}{3}+\frac{2}{3\pi}\operatorname{atan2}\left(\cos\phi, \sin\phi\right), & \nu = -1 \wedge |\phi| > \pi/2\\
		\frac{2}{3}+\frac{1}{3\pi}\operatorname{atan2}\left(\cos\phi, \sin\phi\right), & \nu = -1 \wedge |\phi| \leq \pi/2\\
		\frac{1}{3}+\frac{2}{3\pi}\phi, & \nu = 1
	\end{cases}\\
	a' &= \mathfrak{z}
\end{align}

\subsection{Evaluation of transfer errors}
\label{evalTransferErrors}

One quantitative way to evaluate the ventricular coordinates is to use them to transfer the Euclidean coordinates of a heart $A$ to another heart $B$ and then back again to heart $A$. The Euclidean distance between the original and the transferred Euclidean coordinates can then be computed on heart $A$. This ``two-way error'' was used in~\cite{Bayer-2018-ID11708}.
By ``transferring the Euclidean coordinates of $A$ to $B$'', we mean that we end up with values that can be stored at the discrete mesh nodes of $B$. Therefore, we have to figure out where the point corresponding to a node of $B$ is in $A$. As ventricular coordinates are only precomputed at the nodes of $A$ and $B$, we start with the ventricular coordinates at a node of $B$ and locate the point in $A$ that has the same ventricular coordinates. This point is not necessarily a mesh node of $A$, but is implicitly given through interpolation of ventricular coordinates within the tetrahedron in which it lies. The Euclidean coordinates are then also interpolated from the nodes of $A$ to this point and assigned to the corresponding node of $B$. This process is implemented with the transfer matrix from section~\ref{transferringData} (using linear interpolation).

The two-way error only reflects errors due to non-bijectivity and interpolation.
To capture errors due to inconsistencies in the ventricular coordinates across different geometries, one has to transfer only in one direction, i.e., from heart $A$ to heart $B$, and then compute the deviation to the ground truth on heart $B$. However, no real ground truth is available, because no error-free reference method exists to determine the anatomical point correspondences between both hearts. To overcome this problem, we use the ventricular coordinates themselves to create a synthetic ``mean heart geometry'' for which the ground truth is known by construction. The resulting ``one-way error'' therefore reflects the self-consistency of the ventricular coordinates.

To obtain the novel one-way error between $A$ and $B$, we first transfer the Euclidean coordinates of $B$ to the nodes of $A$. Then, we compute the mean of the original Euclidean coordinates of $A$ and the transferred Euclidean coordinates. Together with the mesh connectivity of $A$, this results in the mean heart geometry $C$. As the node indices of $C$ and $A$ are the same, we can directly copy the ventricular coordinates of $A$ to $C$. These coordinates represent the ground truth. Additionally, a new set of ventricular coordinates is computed on $C$. If the coordinate system is consistent across different geometries, both sets of ventricular coordinates should be the same. To quantify possible inconsistencies, the Euclidean coordinates of $C$ are determined at points where the ventricular coordinates computed on $C$ equal the ventricular coordinates copied from $A$. The Euclidean distance between these and the original Euclidean coordinates is then calculated for each node of $C$. Finally, the one-way error is defined as twice this Euclidean distance, because only half the way between $A$ and $B$ is covered between $A$ and $C$.\\
The one-way error is based on the assumption that the arithmetic mean of two heart geometries again yields a valid heart geometry. This assumption is supported by the widespread use of mean shapes as meaningful and representative geometric objects in the field of statistical shape modeling.

For the test geometries, two- and one-way errors are always computed between the mean shape of the SSM and one of the patient geometries. Both errors are computed for both possible directions, i.e., $A$ and $B$ are interchanged. To obtain the mean geometry for the one-way error, the two geometries to be averaged need to have the same global orientation. For this reason, their heart axes, as determined in section~\ref{heartAxesApex}, are aligned before averaging. After averaging the Euclidean coordinates, the mean geometries have to be remeshed, because the numerical computation of ventricular coordinates requires meshes of sufficient quality, which is not guaranteed after moving the nodes. Linear interpolation is used to transfer the ground truth coordinates onto the remeshed geometries. We use \textit{fTetWild}~\citep{hu2020ftetwild} for remeshing. As an envelope size of $5\,\%$ of the mean edge length is used to approximate the original mesh, the influence of remeshing on the results can be considered negligible.
For further details on the transfer errors, the reader is referred to section~\ref{detailsTransferErrors}.

\subsection{Evaluation of linearity errors}
\label{evalLinearityErrors}

The linearity of coordinates is important to preserve normalized distances. It is particularly relevant for transferring cardiac activation times, where a shortening or lengthening in space can lead to artificial regions of slow or fast conduction, respectively.\\
To evaluate the linearity of the rotational coordinate, we extract contour lines at discrete isovalues of the apicobasal and transmural coordinates and plot the normalized distance along these contour lines over the rotational coordinate. Ideally, the result should be a diagonal line passing through $(0,0)$ and $(1,1)$. The (vertical) absolute deviation from this diagonal is defined as rotational linearity error. The same can be done to obtain an apicobasal linearity error (interchange ``rotational'' and ``apicobasal'' in the previous sentence). To assess the dependency of the rotational (apicobasal) linearity from the apicobasal (rotational) coordinate, we also provide plots of the linearity error over the respective other coordinate. Linearity errors are evaluated separately for both ventricles on all patient geometries. The following isovalues are used for all geometries:
\begin{align}
	a, a' &\in \left\{\tfrac{2}{20}, \tfrac{3}{20}, \dots, \tfrac{19}{20}\right\}\\
	r, r' &\in \left\{\tfrac{1}{72}, \tfrac{3}{72}, \dots, \tfrac{71}{72}\right\}\\
	m, m' &\in \left\{\tfrac{1}{10}, \tfrac{3}{10}, \dots, \tfrac{9}{10}\right\}
\end{align}
This yields 90 contour lines for the evaluation of the rotational and 180 contour lines for the evaluation of the apicobasal linearity error. Linear interpolation was used to resample the normalized distance along the contour lines at 1000 equidistant values of the rotational (apicobasal) coordinate.

%%%%%%%%%%%%%%%%%%%%%%%%%%%%%%%%%%%%%%%%%%%%%%%%%%
\newpage
\section{Results}
\label{results}

\subsection{Computational robustness}
\label{robustness}
Cobiveco was successfully and autonomously computed on all 1000 quasi-random instances of the SSM and all 36 patient geometries, which demonstrates the robustness of the methodology and its implementation.

\subsection{Visual comparison}
\label{visualComparison}
\begin{figure*}[]
	\centering
	\includegraphics[width=\textwidth]{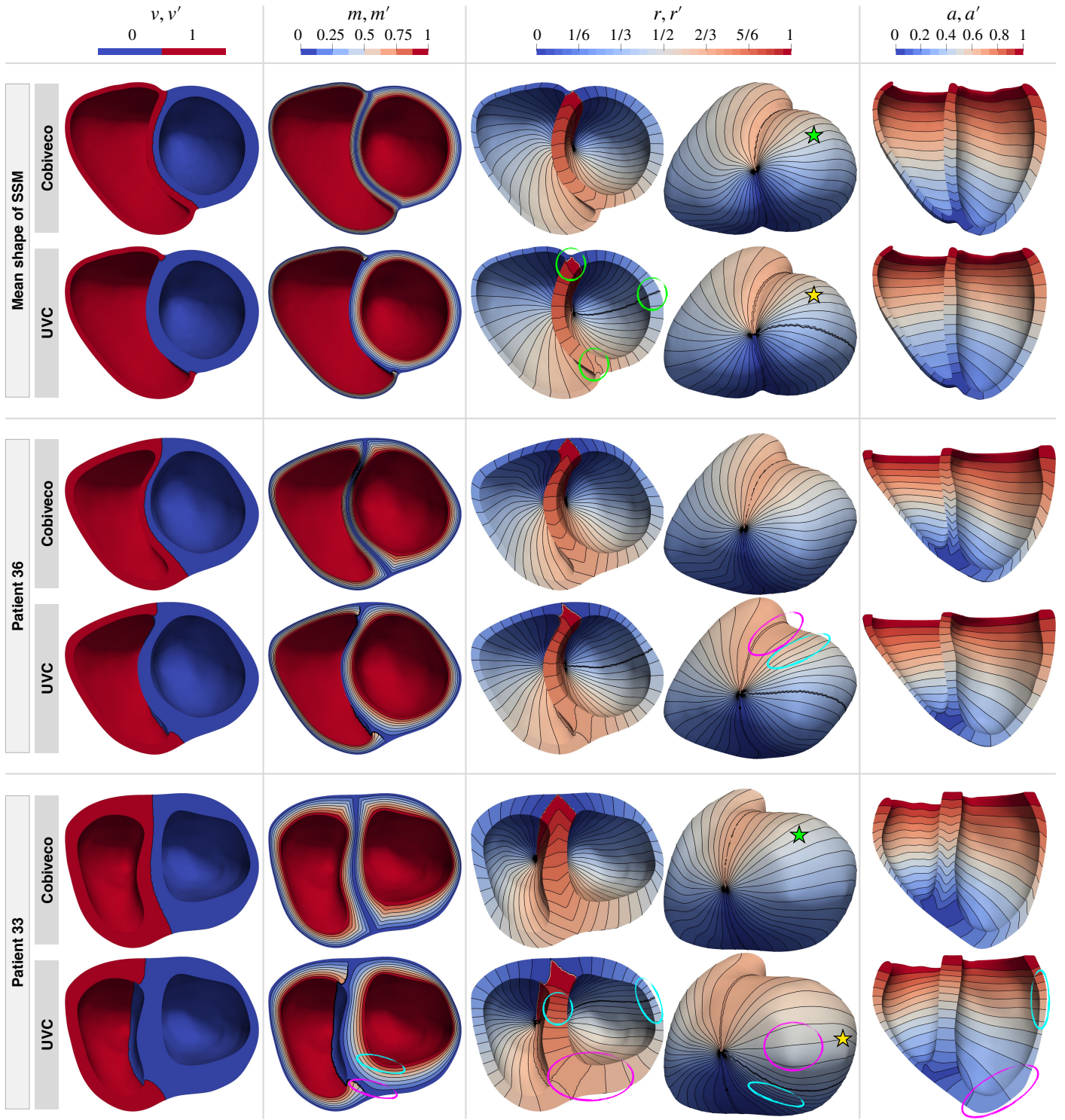}
	\caption{Visual comparison of coordinates $(v,m,r,a)$ and $(v',m',r',a')$ computed using Cobiveco and UVC, respectively, on the mean shape of the SSM and two exemplary patient geometries. Green circles mark artifacts at discontinuities of $r'$. Magenta and cyan circles mark regions of stretched and compressed coordinate values, respectively. Green and yellow stars mark an exemplary point with the coordinates $(0,0,\frac{7}{15},\frac{1}{3})$ for Cobiveco and UVC, respectively.}
	\label{fig:comparison}
\end{figure*}

Fig.~\ref{fig:comparison} provides a visual comparison of Cobiveco and UVC for all four coordinates on the mean shape of the SSM and two exemplary patient geometries.\\
As the mean shape has a very uniform wall thickness, the contour lines of the rotational and apicobasal coordinates appear equidistant for both methods, but artifacts at the discontinuities of the rotational coordinate can be seen for UVC (green circles).\\
Patient 36 also has a relatively uniform wall thickness, but differences between both methods become more apparent. For UVC, the distance between contour lines of the rotational coordinate increases near the septal junctions (magenta vs. cyan circle), which is not the case for Cobiveco.\\
In patient 33, the differences are most pronounced. While the coordinates computed using Cobiveco still change very uniformly in space, there are substantial distortions in the UVC coordinates. The length of the segments between contour lines of the rotational coordinate changes up to four-fold between regions of small and large wall thickness. The apicobasal coordinate is also distributed very non-uniformly, indicating that the geodesic approach to normalize the apicobasal Laplace solution does not work reliably. In fact, a slight change of the geometry can cause a different geodesic path between apex and base to become the shortest and therefore lead to an abrupt change of the apicobasal coordinate. Taking a closer look at the transmural coordinate within the LV shows that it changes much faster at the endocardium than it does at the epicardium because the width of the region between the two boundary surfaces increases with the circumference.\\
If the coordinates always showed the same distortions for every geometry, this would only be a minor problem. However, comparing the rotational and apicobasal UVC coordinates for patient 33 and the mean shape reveals that the same coordinate values can represent quite different anatomical regions (yellow stars). In contrast, the coordinates obtained using Cobiveco are consistent across the geometries (green stars).\\
For pictures showing Cobiveco and UVC coordinates on all 36 patient geometries, the reader is referred to Fig.~\ref{fig:patient_cobiveco}~and~\ref{fig:patient_uvc}, respectively.

\newpage
\subsection{Transfer errors}
\label{transferErrors}
\begin{figure}[!b]
	\begin{minipage}{\textwidth}
			\centering
			\includegraphics[width=\textwidth]{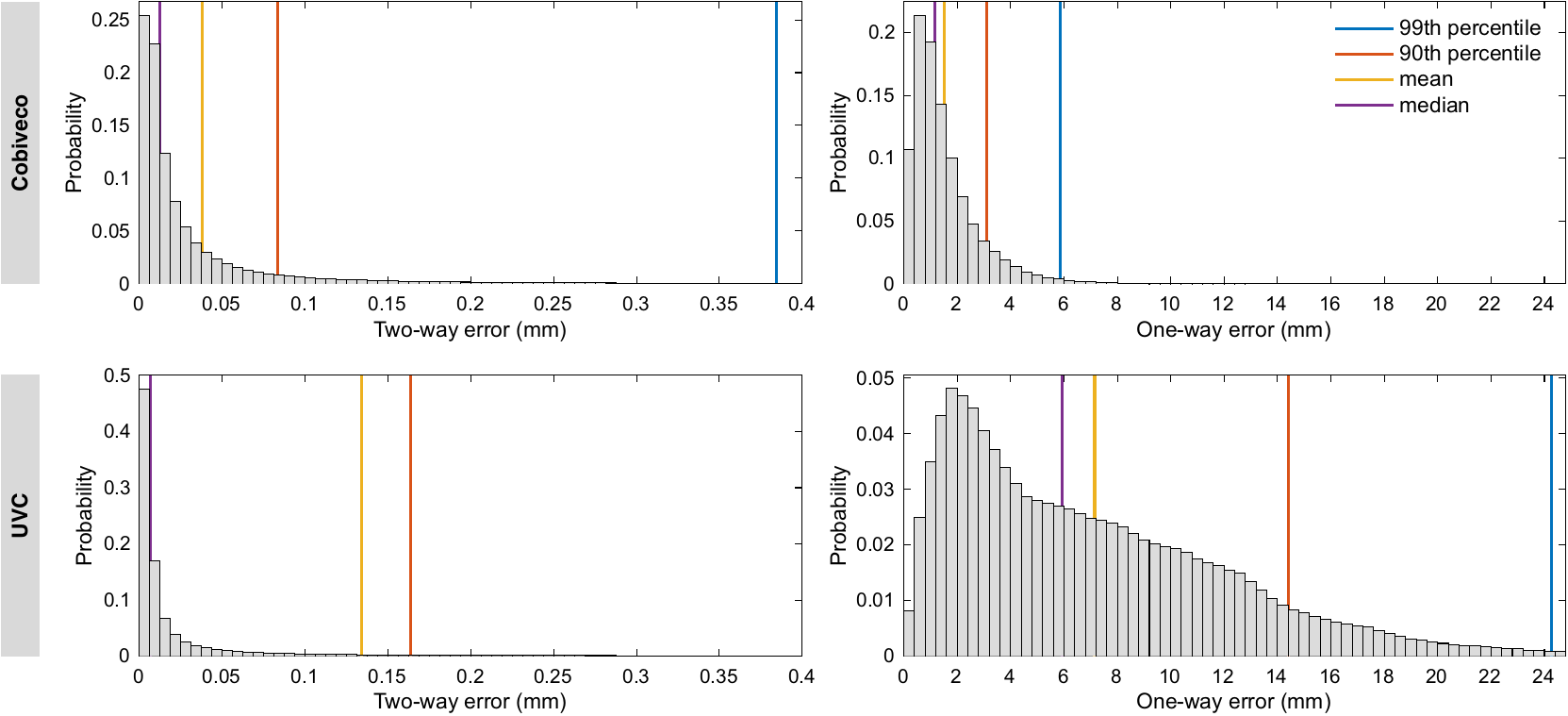}
			\caption{Average histograms of two-way errors (\textit{left}) and one-way errors (\textit{right}) evaluated for Cobiveco (\textit{top}) and UVC (\textit{bottom}). Histograms were averaged across all 36 patients and, for both types of errors, include both possible transfer directions. Each histogram contains about 50\,M data points.}
			\label{fig:transferErrors_histogram}
	\end{minipage}
\end{figure}

\begin{figure*}
	\centering
	\includegraphics[width=\textwidth]{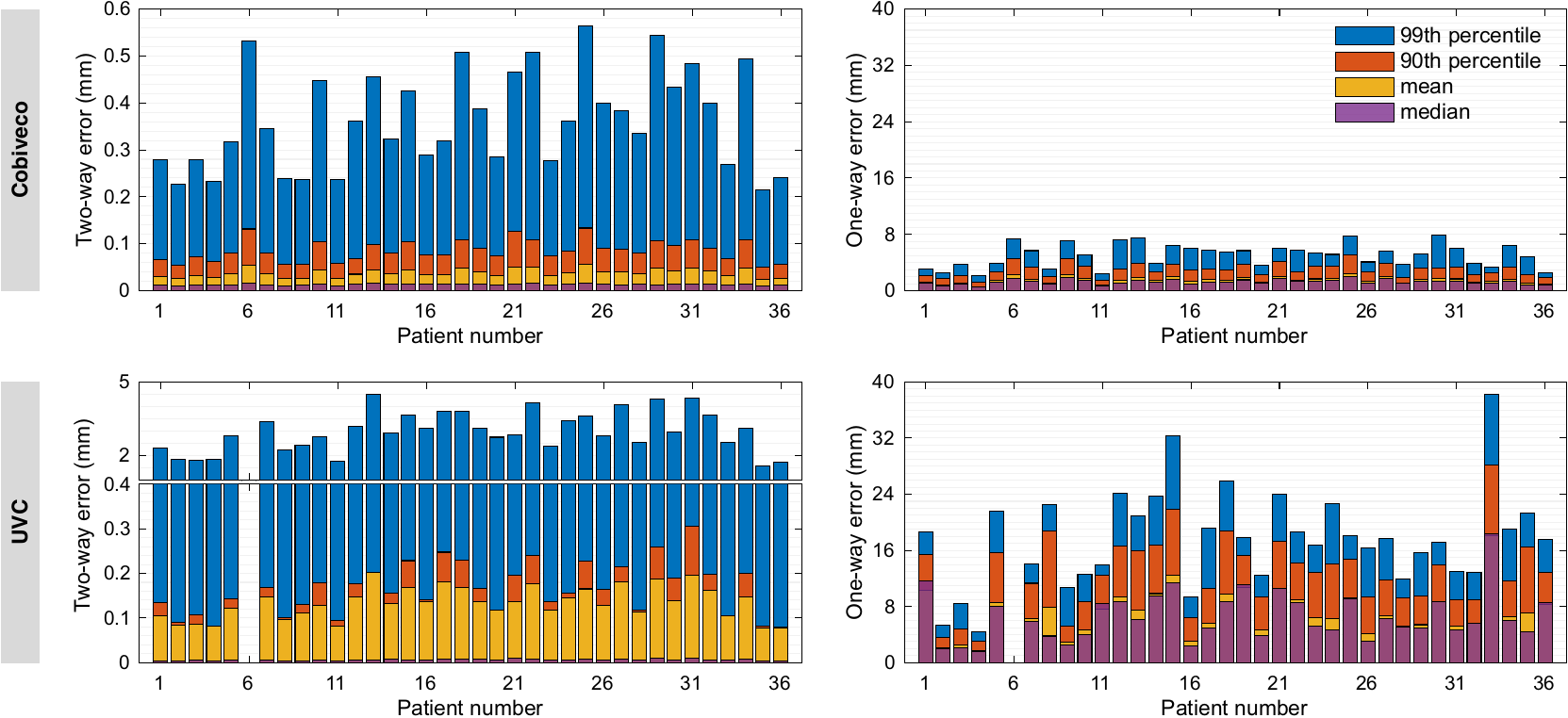}
	\caption{Bar charts showing statistical measures of transfer errors for each individual patient. The data are the same as in Fig.~\ref{fig:transferErrors_histogram}. Patient 6 was excluded from the evaluation of UVC because the rotational and apicobasal UVC coordinates were too inconsistent to obtain a proper mean geometry for the one-way error.}
	\label{fig:transferErrors_modelBars}
\end{figure*}

\begin{figure*}
	\vspace{-2cm}
	\centering
	\includegraphics[width=\textwidth]{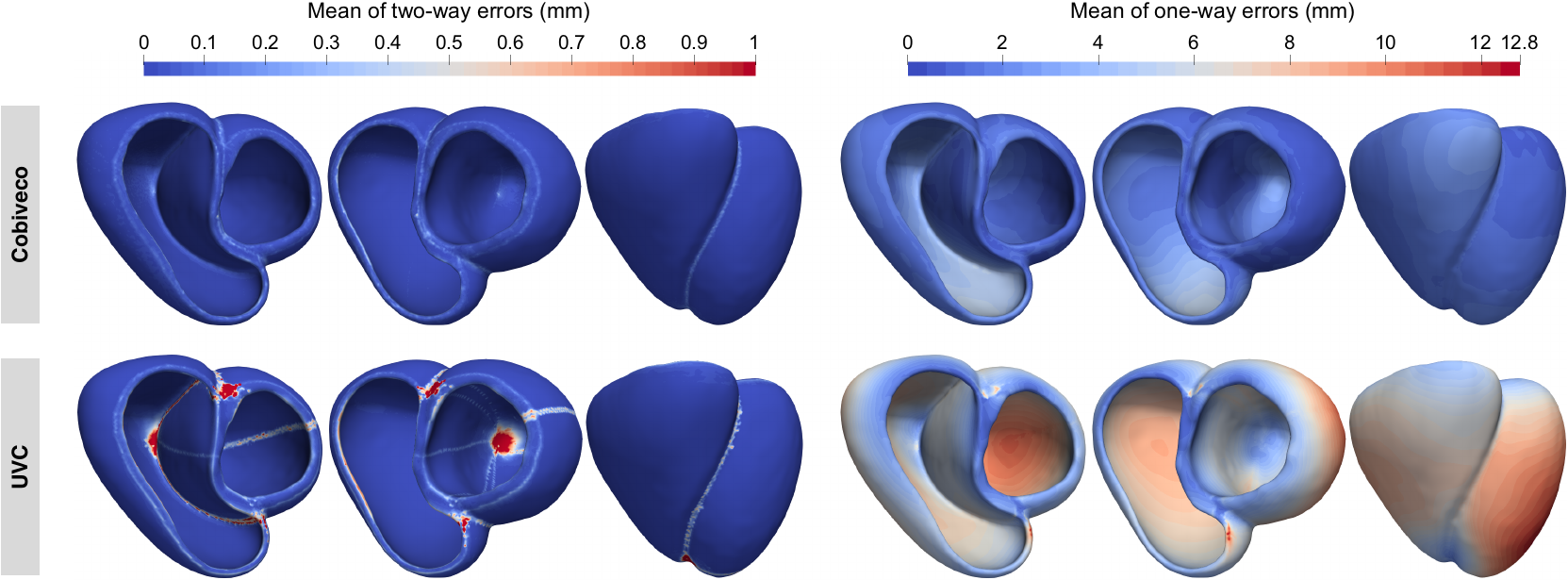}
	\caption{Spatial distribution of the mean transfer errors across all patients visualized on the mean shape of the SSM. As a common geometry is needed to average errors across patients, only transfer sequences with respect to the mean shape of the SSM are included here.}
	\label{fig:transferErrors_meanshape}
\end{figure*}

The transfer errors as defined in section~\ref{evalTransferErrors} were computed for all patient geometries as well as both possible directions. To condense the results, we averaged the error histograms across all geometries and both directions. This leads to an equal weighting of errors for each case, independent of the number of nodes in the respective mesh. The average histograms are depicted in Fig.~\ref{fig:transferErrors_histogram}. Statistical measures (vertical lines) of the average histograms are given in Table~\ref{tab:transferErrors}.\\
The two-way error shows a 3.5-fold improvement of the mean and the 99\textsuperscript{th} percentile is reduced even more. However, the median is increased, which indicates that there are more small (${<}\,0.013\,\mathrm{mm}$), but fewer large errors than for UVC. With a mean value well below one mean edge length, our two-way errors for UVC are comparable to those in~\cite{Bayer-2018-ID11708}.\\
The one-way error is more relevant in practice as it goes beyond evaluation of interpolation errors. Here, the error histogram decays much faster for Cobiveco and all statistical measures show a more than 4-fold improvement compared to UVC. In particular, the mean one-way error is reduced from 7.1 to 1.5\,mm and the 99\textsuperscript{th} percentile is reduced from about 24 to 6\,mm.
\begin{table}[!h]
	\centering
	\vspace{0.5em}
	\caption{Summary of transfer errors (values of the vertical lines in Fig.~\ref{fig:transferErrors_histogram}). All values in mm. Improvements of mean errors are highlighted in bold.}
	\vspace{0.5em}
	\footnotesize
	\setlength{\tabcolsep}{0.45em}
	\newcolumntype{L}[1]{>{\raggedright\let\newline\\\arraybackslash\hspace{0pt}}m{#1}}
	\newcolumntype{R}[1]{>{\raggedleft\let\newline\\\arraybackslash\hspace{0pt}}m{#1}}
	\begin{tabular}{|L{12mm}|L{23mm}|R{9mm}R{9mm}R{9mm}R{9mm}|}
		\hline
		\rowcolor[HTML]{EEEEEE}
		Error type & Coordinate system & Median & Mean & 90\textsuperscript{th} P. & 99\textsuperscript{th} P.\\
		\hline\hline
		\multirow{3}*{Two-way}
		& Cobiveco & 0.013 & 0.038 & 0.084 & 0.385\\
		& UVC & 0.007 & 0.134 & 0.164 & 2.944\\
		& Improvement factor & 0.52 & \textbf{3.52} & 1.96 & 7.66\\
		\hline\hline
		\multirow{3}*{One-way}
		& Cobiveco & 1.17 & 1.51 & 3.10 & 5.87\\ 
		& UVC & 5.93 & 7.15 & 14.4 & 24.26\\
		& Improvement factor & 5.08 & \textbf{4.75} & 4.65 & 4.13\\
		\hline
	\end{tabular}
	\label{tab:transferErrors}
\end{table}
\newpage
\noindent Fig.~\ref{fig:transferErrors_modelBars} shows transfer errors for each individual patient geometry. In all patients, the 99\textsuperscript{th} percentile of the two-way error for Cobiveco is below the mean edge length, which is not the case for UVC (note the broken y-axis). For one-way errors, the largest 99\textsuperscript{th} percentile in a single patient is about 8\,mm for Cobiveco and 38\,mm for UVC.
To assess the spatial distribution of transfer errors, we visualized their mean across all patients on the mean shape of the SSM. To avoid artifacts due to spatial interpolation, only errors directly available on the mean shape of the SSM were taken into account for this purpose, i.e., only one transfer direction was included.
The result in Fig.~\ref{fig:transferErrors_meanshape}~(left) clearly shows that the two-way errors for UVC concentrate at discontinuities of the coordinates (compare with Fig.~\ref{fig:concept}). Furthermore, there are large errors at the singularities of the rotational coordinate. For Cobiveco, these errors are greatly reduced, because only the transventricular coordinate is discontinuous and the origin of the apicobasal coordinate coincides exactly with the rotational singularities. Choosing narrower colormap limits to visualize the two-way errors for Cobiveco (Fig.~\ref{fig:twowayError_meanshape_limited}) reveals the pattern of the isocurves used to compute the apicobasal coordinate (Fig.~\ref{fig:apicobasal}, bottom-left). These many non-zero, but still small errors explain the slight increase in the median two-way error observed for Cobiveco.\\
For the one-way error (Fig.~\ref{fig:transferErrors_meanshape}, right), the discontinuities and singularities only play a minor role. It is dominated by inconsistencies of the coordinates across different geometries, which lead to inconsistent point correspondences. On average, the largest one-way errors occur at the RV outflow tract for Cobiveco and at the apical region of the LV lateral wall for UVC. Nevertheless, absolute errors are much smaller for Cobiveco.


\clearpage
\subsection{Linearity errors}
\label{linearityErrors}
\begin{figure}[!h]
	\begin{minipage}{\textwidth}
		\centering
		\includegraphics[width=\textwidth]{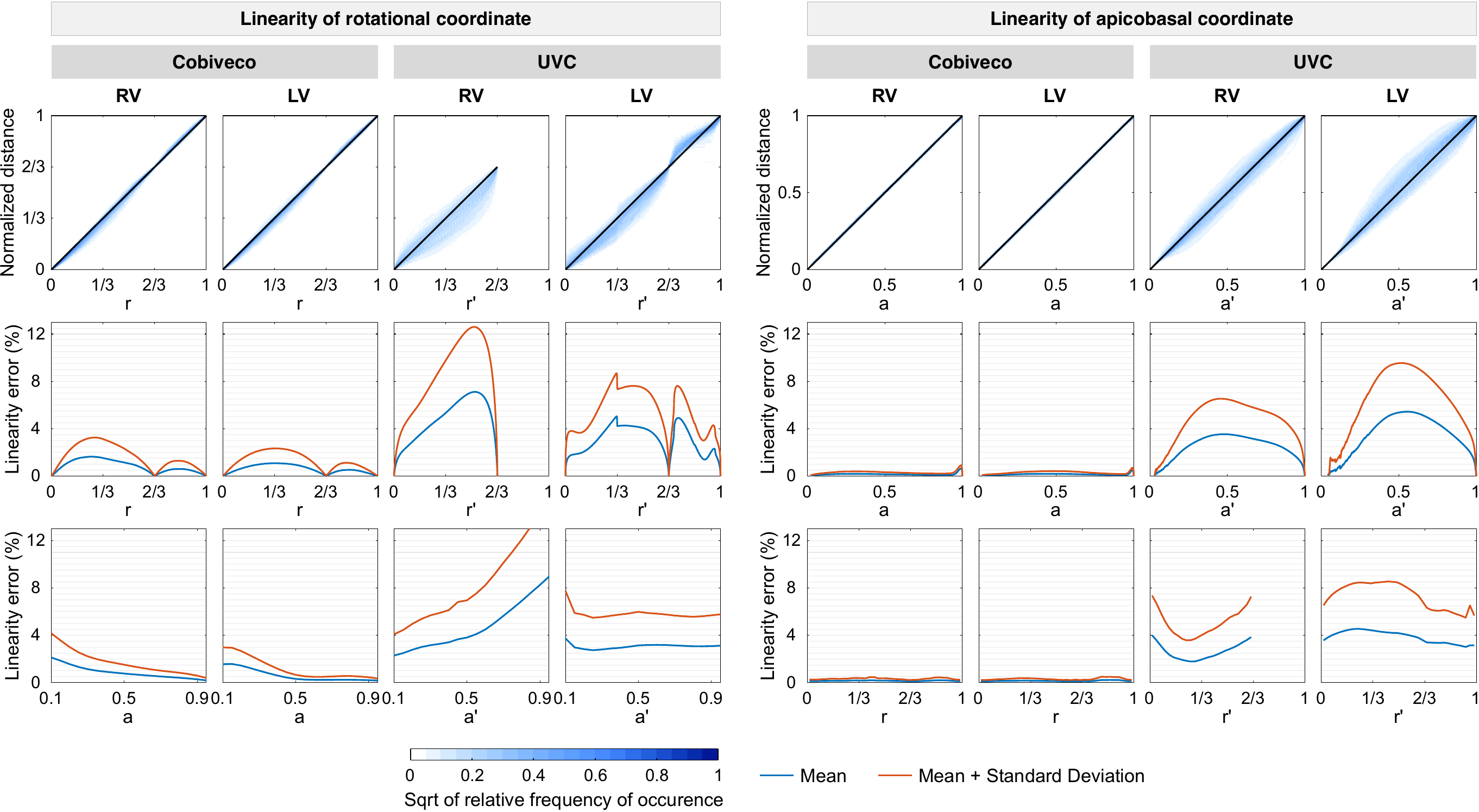}
		\caption{Linearity of the rotational (\textit{left}) and apicobasal (\textit{right}) coordinate of Cobiveco and UVC. In the \textit{first row}, the normalized distance along the contour lines is plotted over the coordinate to be assessed for linearity. A 2D histogram of curves corresponding to the individual contour lines is shown color-coded. Ideally, all points should lie on the black diagonal. The absolute deviation from the black diagonal (linearity error) is plotted in the \textit{second row}. The \textit{third row} shows the dependency of the linearity error on the respective other coordinate (apicobasal coordinate $a$ or $a'$ for the rotational linearity error and rotational coordinate $r$ or $r'$ for the apicobasal linearity error). $r'$ is only defined in the interval $[0,2/3]$ for UVC in the RV.}
		\label{fig:linearityError}
	\end{minipage}
\end{figure}

The linearity of the rotational and apicobasal coordinate was evaluated as described in section~\ref{evalLinearityErrors}. As several thousand contour lines were extracted from the patient geometries, plotting the normalized distance over the respective coordinate for each individual contour line would yield too many curves for visual interpretation. Therefore, we created 2D histograms of these curves. The result can be seen in the first row of Fig.~\ref{fig:linearityError}. The second row shows the actual linearity error, i.e., the absolute deviation from the black diagonal in the first row. Here, the mean and the standard deviation were computed across the different contour lines, but not across the points along a contour line. The third row shows the dependence of the rotational (apicobasal) linearity error on the apicobasal (rotational) coordinate. Here, the mean and the standard deviation were computed along and across all contour lines with the same apicobasal (rotational) isovalue. 
Table~\ref{tab:linearityErrors} summarizes the linearity errors using the maxima of the mean curves in the second row of Fig.~\ref{fig:linearityError}.
\vspace{1.5cm}
\newpage
\noindent In contrast to UVC, the apicobasal coordinate of Cobiveco shows almost perfect linearity. This is expected, as its computation is based on isocontours of the other coordinates.
For both methods, the rotational linearity error is largest in the RV, but Cobiveco also shows a more than 4-fold improvement.
As the circumference of the ventricles increases from apex to base, the (relative) rotational linearity error should decrease with the apicobasal coordinate. This can only be observed for Cobiveco.
\begin{table}[!h]
	\centering
	\caption{Summary of linearity errors (maximum of the curves in the second row of Fig.~\ref{fig:linearityError}). All values in \%. Improvements of mean errors are highlighted in bold.}
	\vspace{0.5em}
	\footnotesize
	\setlength{\tabcolsep}{0.45em}
	\newcolumntype{L}[1]{>{\raggedright\let\newline\\\arraybackslash\hspace{0pt}}m{#1}}
	\newcolumntype{R}[1]{>{\raggedleft\let\newline\\\arraybackslash\hspace{0pt}}m{#1}}
	\begin{tabular}{|L{14mm}|L{23mm}|R{17mm}R{17mm}|}
		\hline
		\rowcolor[HTML]{EEEEEE}
		Coordinate & Coordinate system & RV mean (std) & LV mean (std)\\
		\hline\hline
		\multirow{3}*{\shortstack[l]{Rotational}}
		& Cobiveco & 1.64 (1.59) & 1.09 (1.25)\\
		& UVC & 7.11 (5.48) & 5.04 (3.63)\\
		& Improvement factor & \textbf{4.34} (3.44) & \textbf{4.64} (2.91)\\
		\hline\hline
		\multirow{3}*{\shortstack[l]{Apicobasal}}
		& Cobiveco & 0.65 (0.25) & 0.51 (0.19)\\ 
		& UVC & 3.53 (2.99) & 5.44 (4.05)\\
		& Improvement factor & \textbf{5.42} (11.82) & \textbf{10.58} (21.65)\\
		\hline
	\end{tabular}
	\label{tab:linearityErrors}
\end{table}

%%%%%%%%%%%%%%%%%%%%%%%%%%%%%%%%%%%%%%%%%%%%%%%%%%
\newpage
\section{Application examples}
\label{applicationalExamples}

\subsection{Standardized visualization using polar projections}
\label{polarProjection}
\begin{figure}[!b]
	\centering
	\includegraphics[width=\linewidth]{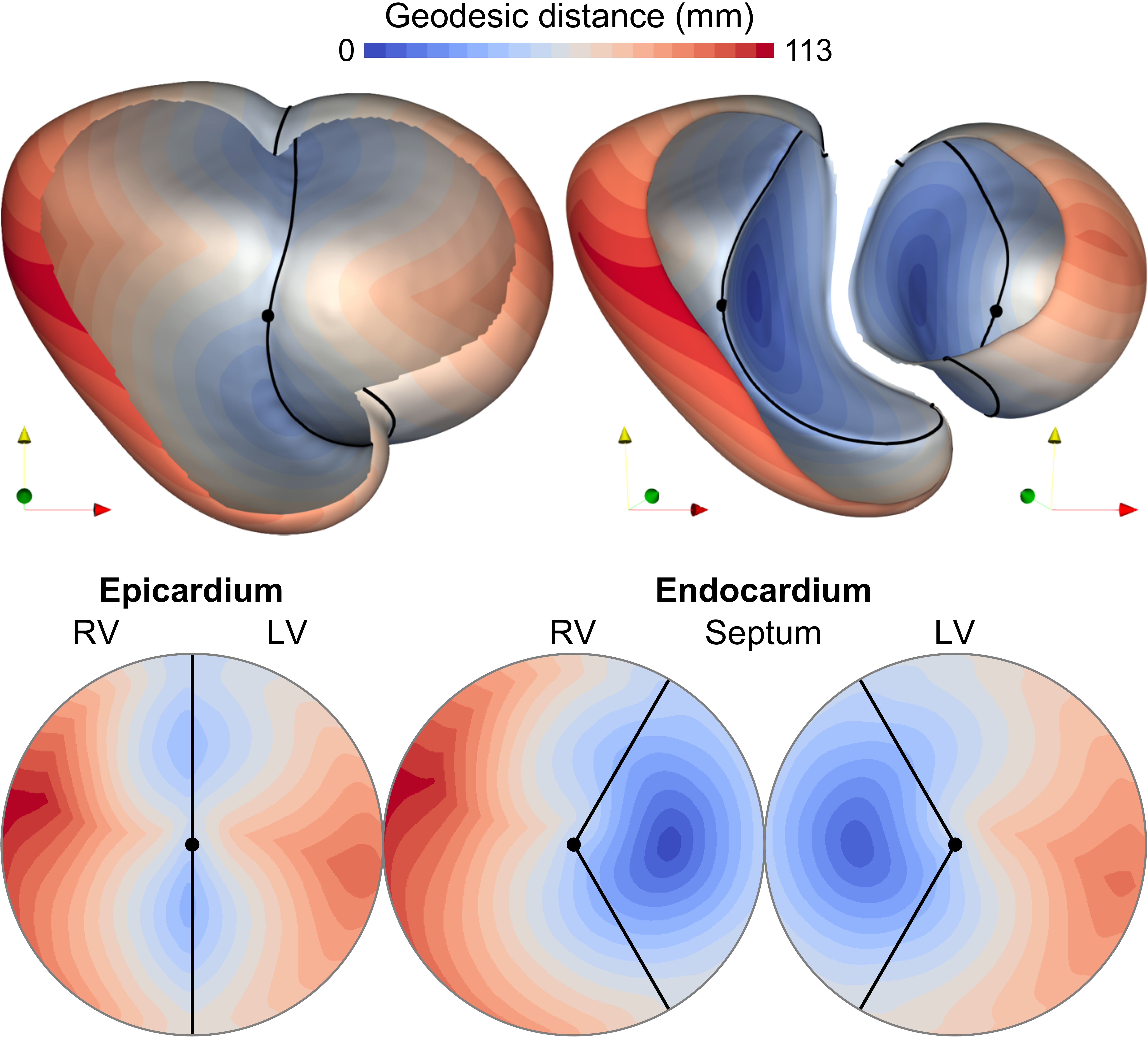}
	\vspace{-1mm}
	\caption{Visualization of a geodesic distance field using polar projections. \textit{Top:} Original data on the epi- and endocardial surfaces. \textit{Bottom}: Corresponding polar plots with projected data. \textit{Black dots}: Apex at $a=0$. \textit{Black lines}: Transventricular/septal junctions at $r=0$ and $r=2/3$.}
	\label{fig:polarProjection}
\end{figure}

One potential application of Cobiveco is the visualization of cardiac data. Apart from the transfer of data from different hearts onto one common biventricular geometry for comparative visualization, the coordinates can also be used for a projection of data onto a 2D representation of the 3D geometry. As a standardized way of visualization, we suggest to represent the surface of the biventricular myocardium using three polar projections: One for the epicardium of both ventricles, one for the RV endocardium, and one for the LV endocardium. Fig.~\ref{fig:polarProjection} shows an example for visualization of a geodesic distance field originating at the center of the RV septal surface. This example was chosen, as it allows a visual assessment of geometric distortions caused by the projections. The main advantage of the polar projections (lower half) is that the entire surface is visible, whereas large regions remain obscured in the corresponding 3D views (upper half) even after individual rotation of the three surfaces.
Polar projections obtained using Cobiveco are an alternative to the method in \cite{stoks2020cinc}, which uses cylindrical coordinates to project a ventricular surface onto a cone and then onto a circular disk and is limited to the epicardium or the LV endocardium only.
Cobiveco also allows to create polar projections for any transmural layer between the endo- and epicardium. To obtain the projections, polar coordinates are computed for a cartesian grid with the desired target resolution. The radial and angular polar coordinates are then interpreted as apicobasal and rotational ventricular coordinates, respectively, and a transfer matrix is constructed as described in section~\ref{transferringData}. We provide a function for computing polar projections as part of the Cobiveco code.

\subsection{Transfer of activation times}
\label{atTransfer}
\begin{figure}[!b]
	\centering
	\includegraphics[width=\linewidth]{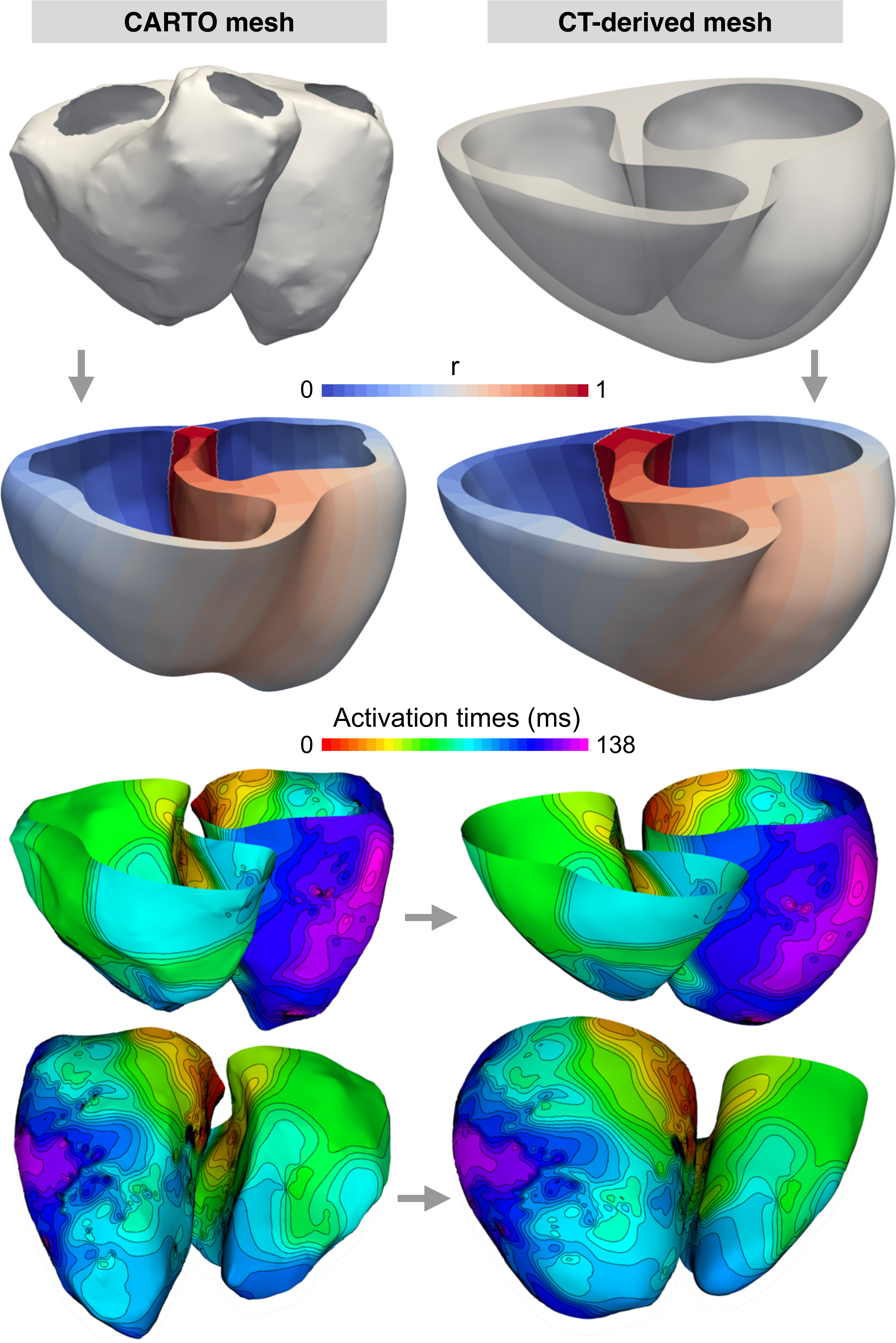}
	\caption{Transfer of activation times recorded using CARTO. \textit{Upper half}: The CARTO mesh is converted into a volume mesh and Cobiveco is computed for both meshes. \textit{Lower half}: The coordinates are used to transfer the activation times from the CARTO mesh to the endocardial surface of the CT-derived mesh.}
	\label{fig:atTransfer}
\end{figure}

Another application is the integration of data from electroanatomical mapping and tomographic imaging.
Fig.~\ref{fig:atTransfer} shows an example for the transfer of activation times recorded using the CARTO mapping system (Biosense Webster, Inc., Irvine, USA) onto the corresponding surfaces of a volume mesh created from CT images. To compute Cobiveco, the endocardial surfaces from CARTO were first converted into a volume mesh (see Fig.~\ref{fig:pipelineVolumeMesh} for a rule-based pipeline to create a volume mesh from only endocardial surfaces). The coordinates obtained on both geometries were then utilized to transfer the activation times.

In contrast to nearest-neighbor mapping~\citep{Duchateau-2018-ID12268,Graham-2019-ID12604} or other straightforward methods~\citep{Cedilnik-2018-ID12306}, Cobiveco allows a continuous and bijective mapping between geometries from both modalities. We believe that an unwanted smoothing of activation times should not motivate a discontinuous mapping between both geometries~\citep{Duchateau-2018-ID12268} but should be addressed by an appropriate spatial upsampling on the source geometry.

%%%%%%%%%%%%%%%%%%%%%%%%%%%%%%%%%%%%%%%%%%%%%%%%%%
\section{Discussion}
\label{discussion}

The evaluation of transfer and linearity errors showed that Cobiveco offers a more consistent description of biventricular position than UVC. These improvements are practically relevant. In the context of ECGI, for example, localization errors lie in the order of 10 to 30\,mm \citep{potyagaylo2019b,graham-2020-ID13306}. The use of ventricular coordinates for validation of non-invasive cardiac mapping or for machine learning based approaches to this problem is only justified if errors due to the coordinates are substantially smaller. Therefore, especially the reduction of the one-way error from 7.1 to 1.5\,mm (mean) and from 24 to 6\,mm (99\textsuperscript{th} percentile) is important.
From the spatial distribution of two- and one-way errors (Fig.~\ref{fig:transferErrors_meanshape}), we conclude that the improvements in continuity of the coordinates help to invert the mapping between ventricular and Euclidean coordinates with high local accuracy (as measured by the two-way error), while the improvements in linearity help to achieve a more consistent mapping between different geometries (as reflected in the one-way error). We put a focus on making the coordinates well-behaved in a mathematical sense, which led us to make the non-binary coordinates continuous also at the LV-RV junction by separating the two ventricles at the center of the septum. Within the septum, the resulting coordinates might not be as intuitive anatomically, but they are intuitive geometrically.
A drawback of Cobiveco compared to UVC is the increased computational complexity. On a modern personal computer (8\,$\times$\,3.8\,GHz CPU), the computation of coordinates for the mean shape of the SSM on a mesh with 479\,k nodes took about 15\,min.
This should be acceptable for the majority of applications, but the efficiency of our implementation could be improved through parallelized remeshing, parallelized isocontour extraction and more advanced preconditioning of linear systems if computational effort becomes crucial for certain use cases. When coordinates have to be obtained for very fine meshes (several millions of nodes), we recommend to increase the sizing parameters of Mmg, which can be passed as an input to Cobiveco. This way, all non-binary coordinates are computed on a coarser mesh and then interpolated to the original mesh. As the coordinates are spatially low-frequent, a mesh resolution of slightly below 1\,mm is sufficient.

\subsection{Limitations}
\label{limitations}
Cobiveco has the following limitations:
\begin{itemize}
	\setlength\itemsep{0em}
	\item The angle between local directions of the apicobasal and the rotational coordinate can become small. This was especially observed near the RV outflow tract (see patient~3 in Fig.~\ref{fig:patient_cobiveco}, for example) and might explain the larger one-way errors in this region (Fig.~\ref{fig:transferErrors_meanshape}, top-right).
	As we decided for normalized and linear coordinates, the angle between coordinate directions directly depends on the shape of the geometry.
	\item The transventricular coordinate remains discontinuous and is defined by a Laplace solution. Although the limitations of Laplace solutions for this purpose are not as severe as for the other coordinates, there might be more accurate ways to separate both ventricles at the center of the septal wall.
	\item For all Laplace solutions, zero Neumann conditions are imposed at boundaries without Dirichlet conditions. This is reasonable for the ridge Laplace solution in \eqref{eq:ridgeLaplace} and for the Laplace solutions used to obtain the tangent fields in \eqref{eq:transmuralTangentField} and \eqref{eq:rotTangentField}. However, natural boundary conditions as suggested in \cite{Stein-2018-ID13150} might be more appropriate for the Laplacian extrapolation in \eqref{eq:lapExtrap} and for the transventricular Laplace solution in \eqref{eq:transventricularLap}.
	\item The coordinate system does not cover myocardial bridges between the atrioventricular valves and the outflow tracts.
\end{itemize}

For the comparisons of UVC and Cobiveco, we reimplemented the UVC method based on the description by \cite{Bayer-2018-ID11708}. We cannot guarantee that this implementation is exactly as the authors of \cite{Bayer-2018-ID11708} intended, although we tried to follow the description in the article and provide our implementation in the Cobiveco repository.

%%%%%%%%%%%%%%%%%%%%%%%%%%%%%%%%%%%%%%%%%%%%%%%%%%
\section{Conclusion}
\label{Conclusion}
We compared different approaches to define and compute ventricular coordinates and developed Cobiveco, a consistent biventricular coordinate system. Key novelties and improvements of Cobiveco are the symmetry of coordinate directions in both ventricles and the definition of coordinate values based on the normalized distance along bijective trajectories between two boundaries, which can be computed by solving linear PDEs. To avoid errors due to imprecise internal boundaries, we use implicit domain remeshing. The resulting coordinates are continuous (when using the sine and cosine transformation for the rotational coordinate and apart from the binary transventricular coordinate), normalized, and change linearly in space. To assess the consistency of the coordinates across different geometries, a novel one-way transfer error was introduced. Evaluation on 36 patient geometries showed a more than 4-fold reduction of transfer and linearity errors compared to UVC. These improvements make Cobiveco an accurate analysis tool and a reliable building block for data-driven modeling of the cardiac ventricles.

%%%%%%%%%%%%%%%%%%%%%%%%%%%%%%%%%%%%%%%%%%%%%%%%%%
\section*{Acknowledgments}
The authors would like to thank Olaf Dössel, Andreas Wachter, Luca Azzolin and Gerald Moik for valuable discussions and feedback. We thank the authors of \cite{Fu-2013-ID14300} for making their implementation\footnote{\url{https://github.com/SCIInstitute/SCI-Solver_Eikonal/tree/v1.0}} of the fast iterative method available to the public.

This work received funding by the German Research Association (DFG) under grants DO 637/21-1 and LO 2093/1-1. This work was supported by the EMPIR programme co-financed by the participating states and from the European Union's Horizon 2020 research and innovation programme under grant MedalCare 18HLT07.

%%%%%%%%%%%%%%%%%%%%%%%%%%%%%%%%%%%%%%%%%%%%%%%%%%
\bibliographystyle{model2-names.bst}\biboptions{authoryear}
\bibliography{ms}

%%%%%%%%%%%%%%%%%%%%%%%%%%%%%%%%%%%%%%%%%%%%%%%%%%%%%%%%%%%%%%%%%%%%%%%%%%%%%%%%%%%%%%%%%%%%%%%%%%%%
\renewcommand{\theHsection}{arabicsection.\thesection}
\renewcommand\thesection{S\arabic{section}}
\setcounter{section}{0}
\renewcommand{\theHfigure}{arabicfigure.\thefigure}
\renewcommand\thefigure{S\arabic{figure}}
\setcounter{figure}{0}
\renewcommand{\theHtable}{arabictable.\thetable}
\renewcommand\thetable{S\arabic{table}}
\setcounter{table}{0}
\renewcommand{\theHequation}{arabicequation.\theequation}
\renewcommand\theequation{S\arabic{equation}}
\setcounter{equation}{0}
\onecolumn
\noindent{\Large Supplementary materials}
\\\ \\

%%%%%%%%%%%%%%%%%%%%%%%%%%%%%%%%%%%%%%%%%%%%%%%%%%
\section{Results on all patient geometries}
\label{resultsAllGeometries}

\begin{figure}[H]
	\centering
	\includegraphics[height=0.99\textheight]{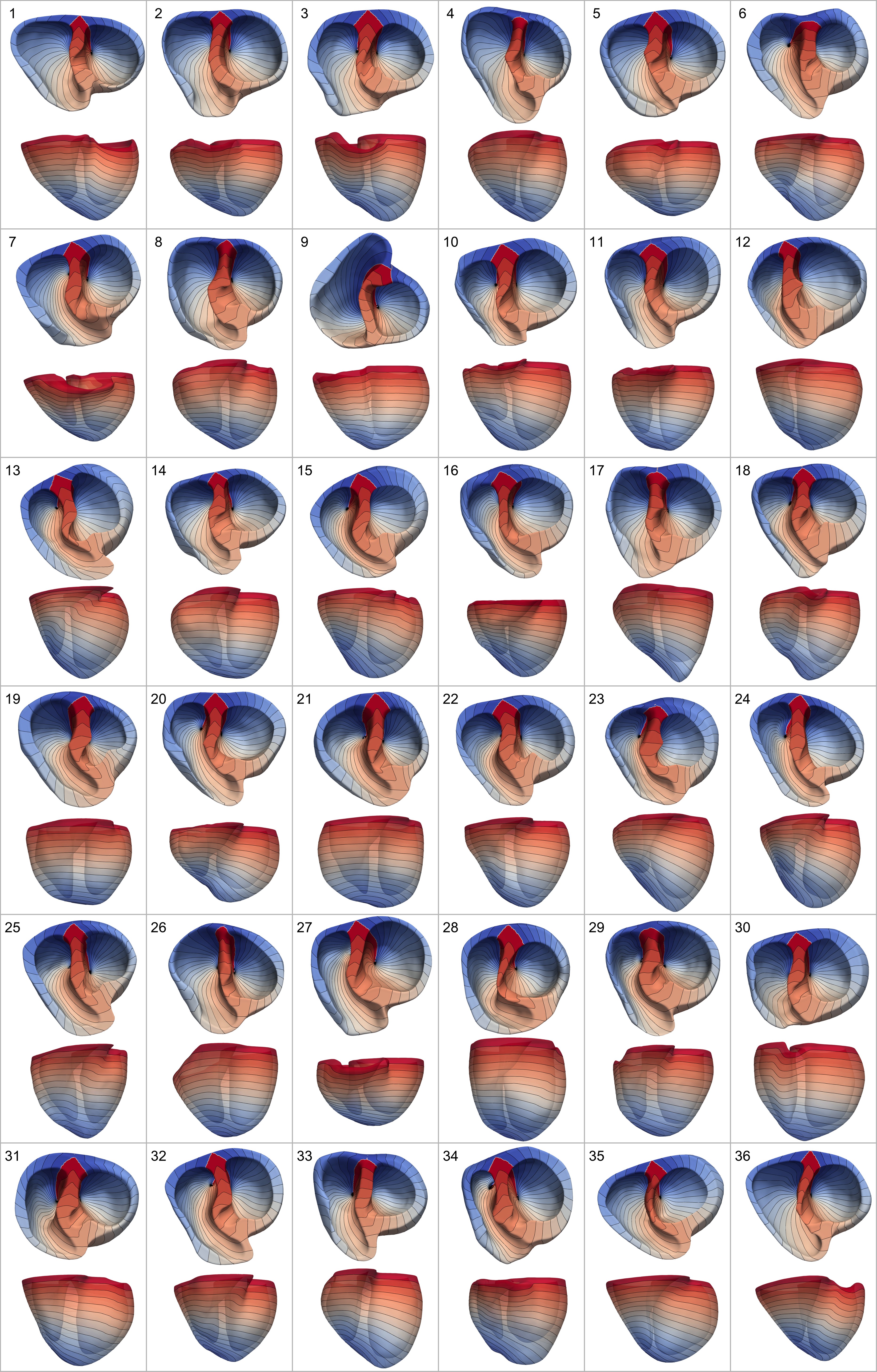}
	\caption{Cobiveco computed for 36 patient geometries. \textit{First rows}: Rotational coordinate $r$. \textit{Second rows}: apicobasal coordinate $a$.}
	\label{fig:patient_cobiveco}
\end{figure}

\begin{figure}[H]
	\centering
	\includegraphics[height=0.99\textheight]{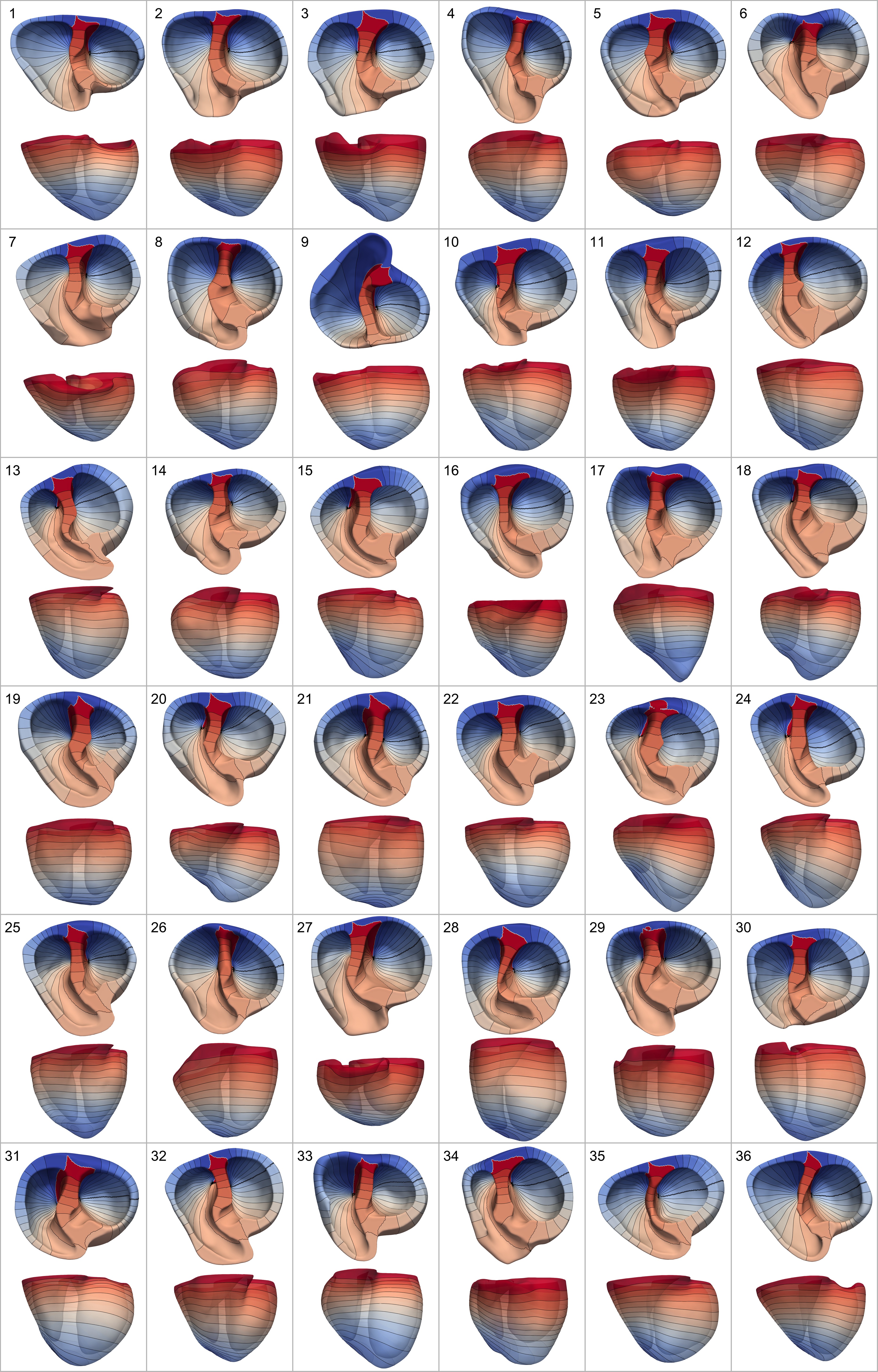}
	\caption{UVC computed for 36 patient geometries. \textit{First rows}: Rotational coordinate $r'$. \textit{Second rows}: apicobasal coordinate $a'$.}
	\label{fig:patient_uvc}
\end{figure}

Table~\ref{tab:meshStatistics} provides mesh statistics for the mean shape of the statistical shape model and the patient geometries. Rotational and apicobasal coordinates computed with Cobiveco on all 36 patient geometries are depicted in Fig.~\ref{fig:patient_cobiveco}. Corresponding UVC coordinates are shown in Fig.~\ref{fig:patient_uvc}.

\begin{table}[H]
	\centering
	\caption{Tetrahedral mesh statistics of the mean shape of the statistical shape model and the 36 patient geometries. $N$: Number of nodes, $\overline{l_e}$: mean edge length, \mbox{$L$: length along the long axis}, $W$: width along the left-right axis. Minimum and maximum values for the patient geometries are marked in bold.}
	\vspace{1em}
	\footnotesize
	\setlength{\tabcolsep}{0.45em}
	\newcolumntype{R}[1]{>{\raggedleft\let\newline\\\arraybackslash\hspace{0pt}}m{#1}}
	\begin{tabular}{|R{7mm}R{6mm}R{6mm}R{6mm}R{6mm}|R{7mm}R{6mm}R{6mm}R{6mm}R{6mm}|}
		\hline
		\rowcolor[HTML]{EEEEEE}
		Mesh (Pat\,\#) & $N$\newline(k) & $\overline{l_e}$\newline(mm) & $L$\newline(cm) & $W$\newline(cm) &
		Mesh (Pat\,\#) & $N$\newline(k) & $\overline{l_e}$\newline(mm) & $L$\newline(cm) & $W$\newline(cm)\\
		\hline\hline
		SSM & 479  & 0.82 & 10.2 & 11.3 & & & & &\\
		\hline
		1   & 900  & 0.66 & 7.9  & 11.5 & 19 & 1103 & 0.91 & 9.9  & 13.5 \\
		2   & 825  & 0.69 & 7.5  & 11.2 & 20 & 1096 & 0.82 & 8.8  & 13.5 \\
		3   & 1029 & 0.71 & 8.6  & 11.1 & 21 & 1216 & 0.96 & 11.5 & 13.9 \\
		4   & 730  & 0.82 & 9.7  & 11.6 & 22 & 1146 & 0.96 & 11.7 & 14.5 \\
		5   & 1105 & 0.93 & 9.5  & 15.3 & 23 & 881  & 0.86 & 11.1 & 12.2 \\
		6   & \textbf{1425} & 0.75 & 9.4  & 12.7 & 24 & 953  & 0.98 & 12.2 & 14.3 \\
		7   & 1074 & 0.79 & 8.4  & 13.6 & 25 & 1092 & 0.81 & 10.5 & 11.8 \\
		8   & 903  & 0.70 & 8.1  & 10.2 & 26 & 885  & 0.86 & 10.5 & 12.8 \\
		9   & 781  & 0.67 & 7.3  & 10.7 & 27 & 1146 & 0.78 & 7.9  & 12.4 \\
		10  & 1032 & 0.92 & 11.0 & 14.0 & 28 & 848  & 1.04 & 12.9 & 13.1 \\
		11  & 938  & \textbf{0.58} & \textbf{7.0}  & \textbf{8.6}  & 29 & 1327 & 0.76 & 9.4  & 11.4 \\
		12  & 872  & 1.01 & 11.8 & 14.7 & 30 & 934  & 1.01 & 12.1 & 14.8 \\
		13  & 912  & \textbf{1.07} & \textbf{13.0} & \textbf{15.4} & 31 & 1283 & 0.80 & 9.6  & 12.4 \\
		14  & 977  & 0.94 & 11.2 & 14.4 & 32 & 959  & 0.83 & 9.6  & 12.4 \\
		15  & 1105 & 0.83 & 9.9  & 12.7 & 33 & 972  & 0.81 & 10.4 & 11.6 \\
		16  & 862  & 0.88 & 10.3 & 12.8 & 34 & 1391 & 0.68 & 8.4  & 9.8  \\
		17  & 924  & 0.82 & 10.7 & 12.1 & 35 & 703  & 0.75 & 8.4  & 11.9 \\
		18  & 1268 & 0.81 & 10.3 & 12.2 & 36 & \textbf{638}  & 0.90 & 10.4 & 14.6 \\
		\hline
	\end{tabular}
	\label{tab:meshStatistics}
\end{table}

\newpage

%%%%%%%%%%%%%%%%%%%%%%%%%%%%%%%%%%%%%%%%%%%%%%%%%%
\section{Details on the transfer errors}
\label{detailsTransferErrors}

\subsection{Mathematical description of two- and one-way errors}
Let $\mathbf{X}_A \in \mathbb{R}^{N_A\times 3}$ and $\mathbf{X}_B \in \mathbb{R}^{N_B\times 3}$ denote the Euclidean coordinates of the hearts~$A$ and $B$, respectively. The ventricular coordinates computed on these hearts are denoted $\mathbf{V}_A \in \mathbb{R}^{N_A\times 4}$ and $\mathbf{V}_B \in \mathbb{R}^{N_B\times 4}$. $N_A$ and $N_B$ are the numbers of nodes.
Transferring the Euclidean coordinates of $A$ to the nodes of $B$ can be expressed as multiplication with the transfer matrix $\mathbf{M}_{B\leftarrow A} \in \mathbb{R}^{N_B\times N_A}$:
\begin{equation}
	\mathbf{M}_{B\leftarrow A}\,\mathbf{X}_A \label{eq:transferEuclidean}
\end{equation}
$\mathbf{M}_{B\leftarrow A}$ is computed as described in section~2.5 (using linear interpolation). It depends on the ventricular coordinates of $A$ and $B$ (and the mesh connectivity of $A$) but not on their Euclidean coordinates:
\begin{equation}
	\mathbf{M}_{B\leftarrow A} = f(\mathbf{V}_A,\mathbf{V}_B)
\end{equation}

Using the notation from~\eqref{eq:transferEuclidean}, the two-way error for the transfer sequence ``$A$ to $B$ and back to $A$'' can be written as:
\begin{equation}
	\mathbf{e}_{AB}^\text{two-way} = \|\mathbf{X}_A - \widetilde{\mathbf{X}}_A\|_\mathrm{col} \quad\text{with}\quad \widetilde{\mathbf{X}}_A = \mathbf{M}_{A\leftarrow B}\,\mathbf{M}_{B\leftarrow A}\,\mathbf{X}_A
	\label{eq:twoway}
\end{equation}
Here, $\|\cdot\|_\mathrm{col}$ denotes the $2$-norm along the column dimension.\\

The one-way error between $A$ and $B$ with respect to $A$ is defined as:
\begin{equation}
	\mathbf{e}_{AB}^\text{one-way} = 2\,\|\mathbf{X}_C - \widetilde{\mathbf{X}}_C\|_\mathrm{col}
	\label{eq:oneway}
\end{equation}
with
\begin{align}
	\mathbf{X}_C &= \tfrac{1}{2}(\mathbf{X}_A+\mathbf{M}_{A\leftarrow B}\,\mathbf{X}_B) \label{eq:onewayXC}\\
	\widetilde{\mathbf{X}}_C &= \mathbf{M}_{A\leftarrow C}\,\mathbf{X}_C \label{eq:onewayXtilde}
\end{align}
$\mathbf{X}_C$ is the average of the Euclidean coordinates at the nodes of~$A$ and the Euclidean coordinates at the corresponding points in~$B$. Together with the mesh connectivity of $A$, it forms the mean heart geometry $C$. As the node indices of $A$ and $C$ are identical, we can directly copy the ventricular coordinates computed on $A$ to $C$. This yields the ``ground truth'' coordinates $\mathbf{V}_A$ on $C$. A new set of coordinates ``to be evaluated'' $\mathbf{V}_C$ is then computed on $C$. $\mathbf{M}_{A\leftarrow C}\,\mathbf{X}_C$ yields the Euclidean coordinates of $C$ at points where $\mathbf{V}_C$ equals $\mathbf{V}_A$. If the coordinate system is consistent across different geometries, $\mathbf{M}_{A\leftarrow C}$ should be close to the identity matrix and the norm of the difference in \eqref{eq:oneway} should be small. As the transfer between $A$ and $C$ covers only half the way between $A$ and $B$, the norm in \eqref{eq:oneway} is multiplied by two.

\subsection{Illustration of the one-way error}
\begin{figure*}[h!]
	\centering
	\includegraphics[width=\textwidth]{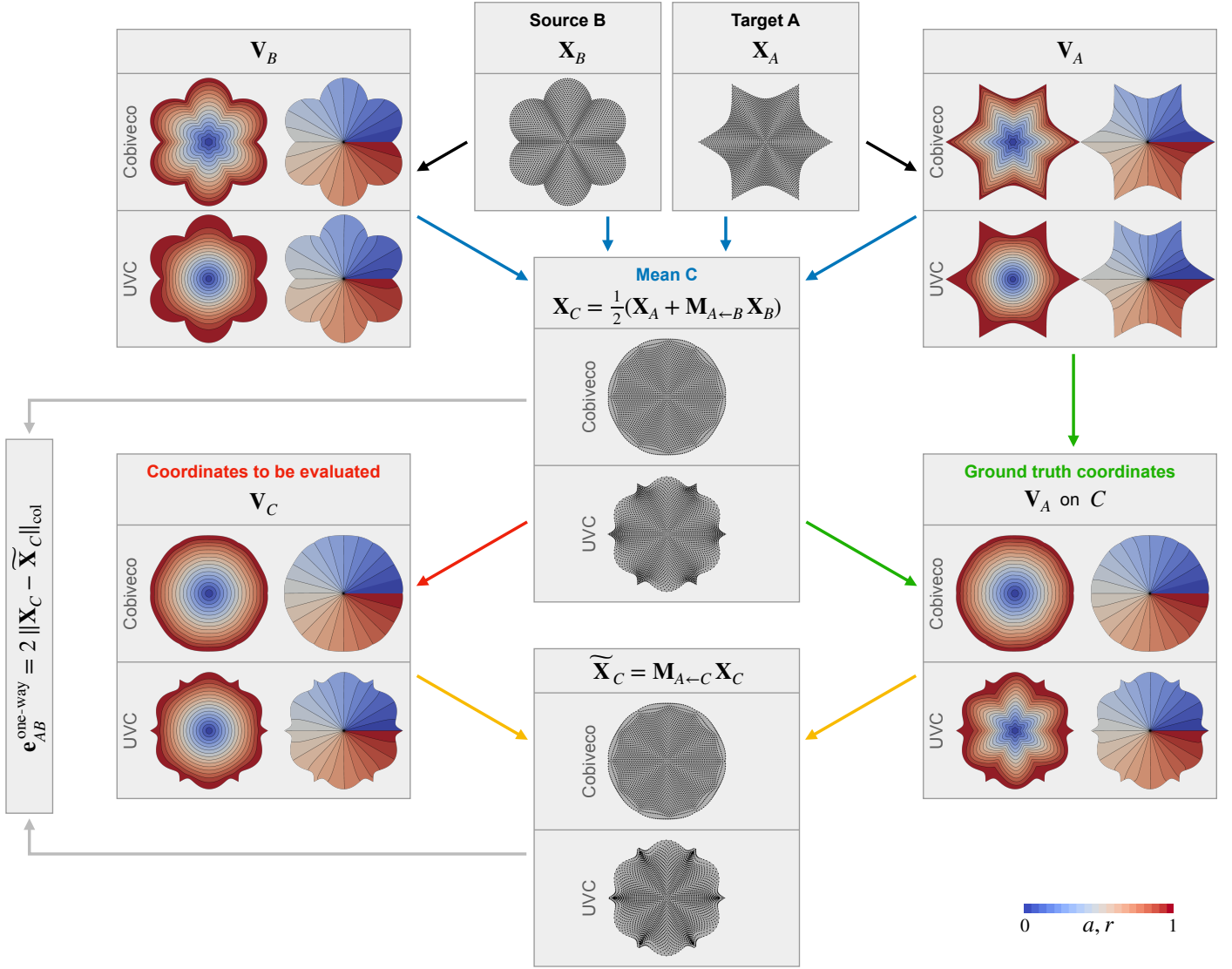}
	\caption{Two-dimensional example illustrating the computation of the one-way error in \eqref{eq:oneway}--\eqref{eq:onewayXtilde}. We start with the Euclidean coordinates $\mathbf{X}_B$ and $\mathbf{X}_A$ of a source geometry $B$ and a target geometry $A$, for which we compute the ventricular coordinates $\mathbf{V}_B$ and $\mathbf{V}_A$ (black arrows). These ventricular coordinates are used to map $\mathbf{X}_B$ to $\mathbf{X}_A$ and the result is averaged with $\mathbf{X}_A$, yielding a mean geometry $C$ represented by $\mathbf{X}_C$ (blue arrows). For this geometry, ventricular coordinates $\mathbf{V}_C$ can be calculated (red arrow). As $A$ and $C$ have the same node indices, we can furthermore copy $\mathbf{V}_A$ to $C$ to obtain ground truth coordinates on $C$ (green arrows). The Euclidean coordinates of $C$ are then determined at points where $\mathbf{V}_C$ equals $\mathbf{V}_A$, which yields $\widetilde{\mathbf{X}}_C$ (yellow arrows). Finally, $\mathbf{X}_C$ is compared with $\widetilde{\mathbf{X}}_C$ (gray arrows).}
	\label{fig:onewayError}
\end{figure*}

Fig.~\ref{fig:onewayError} illustrates the principle of the one-way error using a simple two-dimensional example. Here, the hearts $A$ and $B$ are represented by a star-shaped and a flower-shaped geometry, respectively. For both geometries, ``ventricular'' coordinates are computed analogously to the definition of the apicobasal and the rotational coordinates in Cobiveco and UVC (black arrows). In this example, averaging both geometries using the Cobiveco coordinates yields an almost circular mean geometry, while the UVC coordinates lead to a spiky mean geometry (blue arrows). Two sets of ventricular coordinates are then obtained on the respective mean geometry: The coordinates to be evaluated are computed independently (red arrow), while the ground truth coordinates are copied from~$A$ (green arrows). For Cobiveco, these two sets of ventricular coordinates look very similar, while larger differences can be seen for UVC, especially for the apicobasal coordinate. To quantify the Euclidean distance associated with this inconsistency of ventricular coordinates, the Euclidean coordinates of $C$ are determined at points where the coordinates to be evaluated equal the ground truth coordinates (yellow arrows). The result is then compared with the Euclidean coordinates of the mean geometry (gray arrows).

\begin{figure}[h!]
	\centering
	\includegraphics[width=\textwidth]{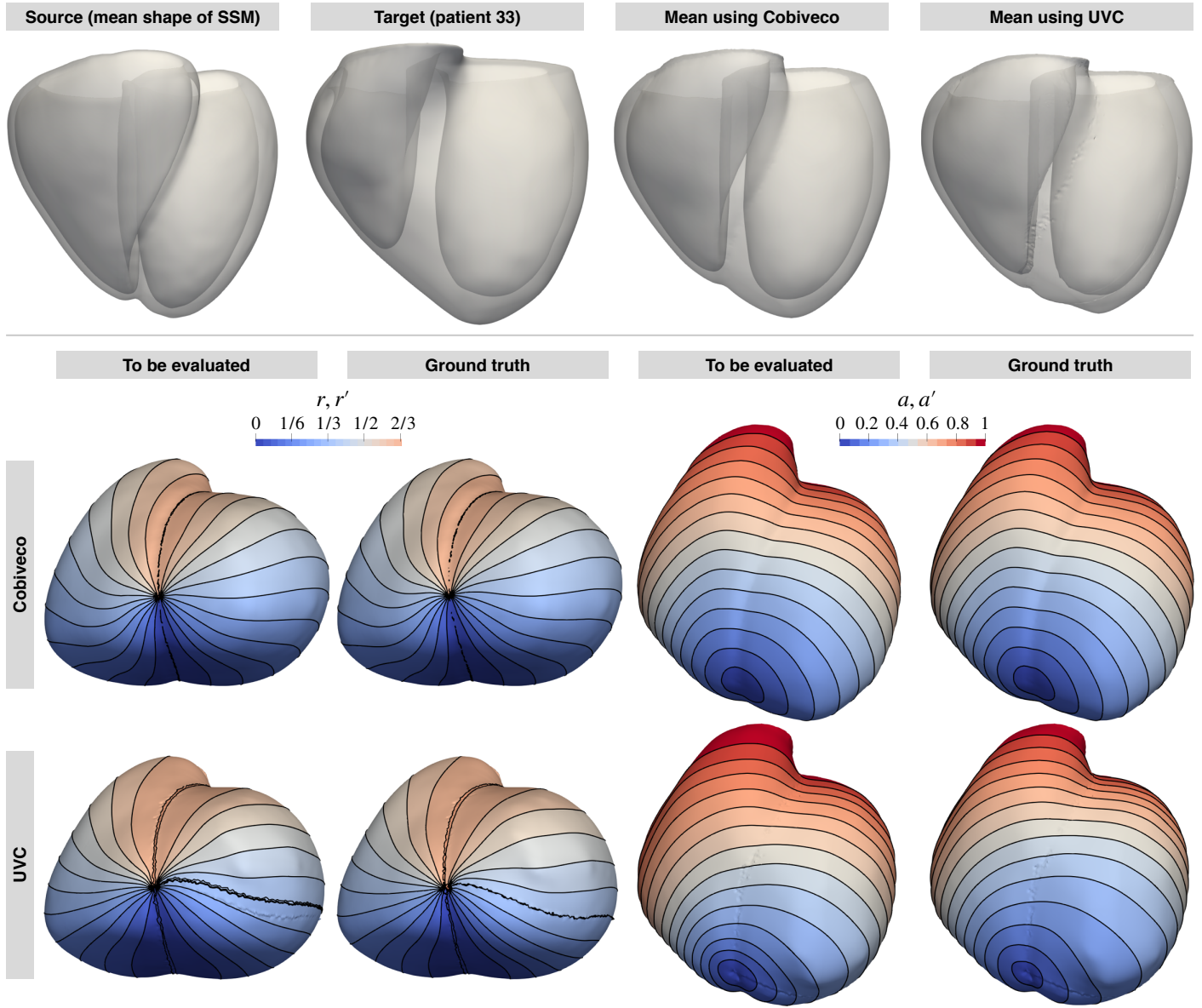}
	\caption{Geometries and coordinates involved in the computation of the one-way error for patient 33. \textit{Upper panel}: Source, target and mean geometries. \textit{Lower panel}: Coordinates to be evaluated and ground truth coordinates on the mean geometry.}
	\label{fig:onewayError_pat33}
\end{figure}

Fig.~\ref{fig:onewayError_pat33} gives an example of the actual geometries and coordinates involved in the computation of the one-way error.
\clearpage

\clearpage

\subsection{Additional visualization of two-way errors}
\begin{figure}[H]
	\centering
	\includegraphics[width=0.4\textwidth]{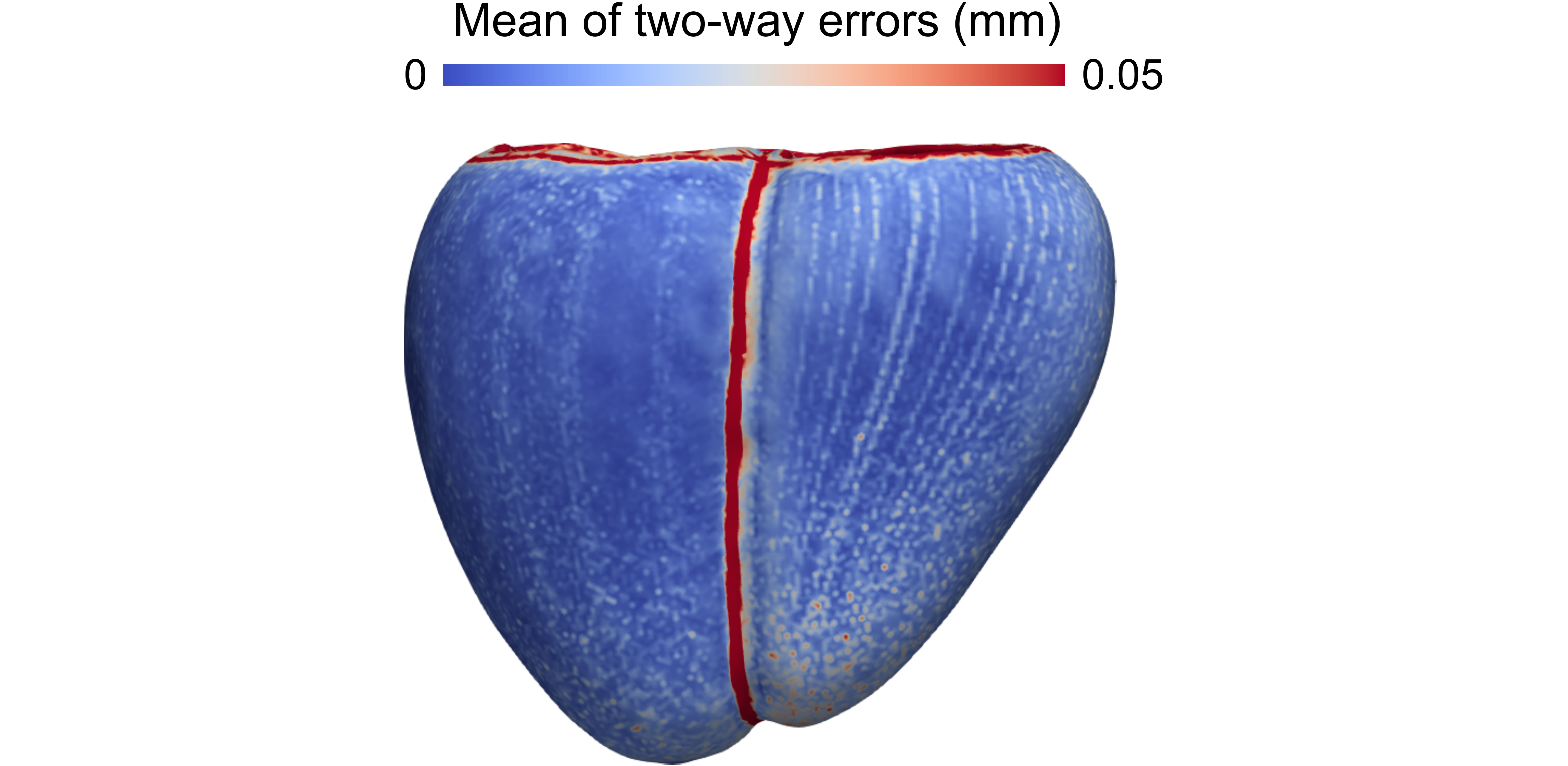}
	\vspace{2mm}
	\caption{Visualization of the mean two-way errors for Cobiveco as in Fig.~\ref{fig:transferErrors_meanshape}, but with narrower colormap limits. The pattern of isocurves used to compute the apicobasal coordinate is visible.}
	\label{fig:twowayError_meanshape_limited}
\end{figure}

Fig.~\ref{fig:twowayError_meanshape_limited} shows the mean two-way errors resulting for Cobiveco. The colormap limits were narrowed compared to Fig.~17, which reveals the pattern of isocurves used to compute the apicobasal coordinate.

%%%%%%%%%%%%%%%%%%%%%%%%%%%%%%%%%%%%%%%%%%%%%%%%%%
\section{Rule-based pipeline to create a volume mesh from endocardial surfaces}
\label{pipelineVolumeMesh}
\begin{figure}[H]
	\centering
	\includegraphics[width=\textwidth]{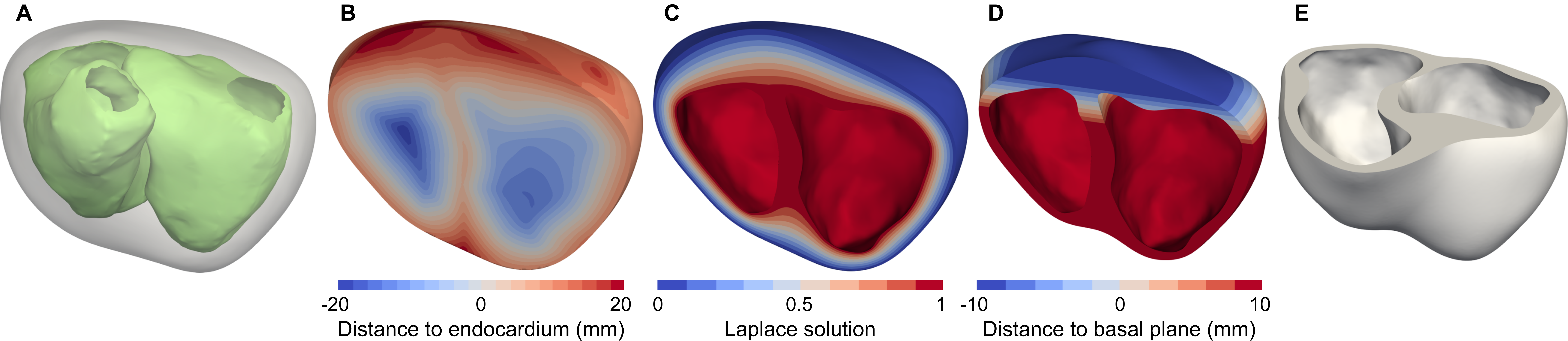}
	\caption{Pipeline for creating a biventricular volume mesh from endocardial surfaces of the LV and RV. \textit{A:} A dilated convex hull (gray) is computed for the endocardial surfaces (green). \textit{B:} A background mesh is created by tetrahedralizing the convex hull and the signed Euclidean distance to the endocardial surfaces is computed for each node. \textit{C:} The mesh region outside the endocardial surfaces is extracted by applying isovalue discretization at 0 to \textit{B} and a Laplace solution between the inner (1) and outer (0) surfaces is computed. \textit{D:} Isovalue discretization at 0.5 is applied to \textit{C} and the distance to a plane defining the base is computed. \textit{E:} Isovalue discretization at 0 is applied to \textit{D}, which yields the final volume mesh. \textit{B}--\textit{D} are clipped for visualization.}
	\label{fig:pipelineVolumeMesh}
\end{figure}

Fig.~\ref{fig:pipelineVolumeMesh} illustrates a rule-based pipeline to create a volume mesh from endocardial surfaces, which makes it possible to compute Cobiveco for geometries for which only endocardial surfaces are available.

\end{document}